\documentclass[fleqn,usenatbib]{mnras}

\usepackage{newtxtext,newtxmath}

\usepackage[T1]{fontenc}

\DeclareRobustCommand{\VAN}[3]{#2}
\let\VANthebibliography\thebibliography
\def\thebibliography{\DeclareRobustCommand{\VAN}[3]{##3}\VANthebibliography}

\usepackage{graphicx}	
\usepackage{amsmath}	
\usepackage{color}

\newcommand{\heii}{He\,{\sc II}}

\title[\textsc{X-BPASS}]{\textsc{X-BPASS} : Self-consistent modelling of stellar populations and their associated X-ray Binary emission in a binary stellar evolution framework}

\author[J. C. Bray. E. R. Stanway J.J. Eldridge]{
J. C. Bray,$^{1}$\thanks{E-mail: john.bray@auckland.ac.nz}
E. R. Stanway,$^{2}$,
J. J. Eldridge$^{1}$
\\
$^{1}$Department of Physics, University of Auckland, Private Bag 92019, Auckland, New Zealand\\
$^{2}$Department of Physics, University of Warwick, Gibbet Hill Road, Coventry, CV5 8EP, United Kingdom\\
}

\date{Accepted XXX. Received YYY; in original form ZZZ}

\pubyear{2015}

\begin{document}
\label{firstpage}
\pagerange{\pageref{firstpage}--\pageref{lastpage}}
\maketitle

\begin{abstract}
X-ray binaries play a significant role in the thermal and ionization history of galaxies. Their X-ray luminosity can shed light on galactic star formation rates and histories. Compact objects are also crucial in the evolution of gravitational wave progenitors. Here we present the results from our work to extend the binary population and spectral synthesis (\textsc{BPASS}) code suite to incorporate X-ray emission onto compact remnants in binary systems. We self-consistently model the accretion disc for each interacting binary system in a grid of stellar evolution models and then combine these to obtain the total X-ray spectra for stellar populations over a range of ages and metallicities. Crucially, these are estimated using the same stellar models as those used for modelling the stellar spectral energy distribution. We utilise first principle equations to calculate the X-ray binary (XRB) evolution, luminosity and spectral energy densities of individual accreting compact objects. 
Population synthesis using observationally motivated values for $R_{\textrm{inner}}$ (the accretion disc inner truncation radius) reproduces the observed X-ray number evolution in the Small Magellanic Cloud and the inferred X-ray flux evolution  for M51, validating our models. 
Using these models, we explore the implications of a self-consistent stellar and XRB emission population synthesis for ionizing photon production, the XRB dependence on metallicity and, for XRBs as a potential source of nebular \heii\ emission seen in the spectra of high-redshift galaxies. 
We conclude that XRBs contribute towards powering nebular \heii\ emission without causing significant overestimates of hydrogen ionization.
\end{abstract}

\begin{keywords}
Stars : binaries : close -- stars : black holes -- stars : neutron -- stars : white dwarf -- stars : massive -- X-rays : binaries -- X-rays : galaxies : clusters 
\end{keywords}



\section{Introduction}



It has been suggested that high mass X-ray binaries (HMXBs), play a significant role in the thermal and ionisation history of galaxies \citep{Bluem2019, Brorby2016, Fragos2013a, Fragos2013b}. 
The X-ray luminosity of a stellar population or galaxy can also be used to understand its star-formation rate and/or history \citep{Yang2022, Brorby2016, Fragos2013a, Fragos2013b, Renalli2003}. However, population modelling of X-ray binaries (XRBs), is often undertaken using simple prescriptions, based on assumptions regarding the initial stellar mass function and the remnant masses produced by massive stars. More detailed analysis on XRB and compact remnant populations such as \cite{Rocha2024, Garofali2023, Liu2023,Fragos2009,Podsiadlowski_2003} has been carried out, albeit without consideration of the simultaneous stellar emission. In this work we aim to combine the detailed \textsc{BPASS} stellar models with aspects of the more detailed analyses of the accretion process in the research above to model the XRB emission in conjunction with the accompanying stellar emission. A similar approach was used by \cite{Lecroq2023}, but since their work was published after the submission of this paper, we have not carried out a detailed comparison to their results.

The Binary Population and Spectral Synthesis (\textsc{BPASS}) project has been used to study many aspects of stellar populations, from our own Galaxy to the edge of the observable Universe. \textsc{BPASS} is also widely used to understand supernovae, their progenitors and gravitational wave transients \citep[e.g.][]{Stanway2016, Eldridgeetal2017, 2017ApJ...848L..28L, StanwayEldridge2018, Eldridgeetal2019, 2022MNRAS.513.3550C, 2023ApJ...955L..29P}. Perhaps most significantly it has been used to show how important the inclusion of interacting binary stars are at the earliest times of the Universe to aid re-ionization 
\citep[e.g.][]{Stanway2016,Wilkins2016,Ma2016,Rosdahl2018} and in the interpretation of spectral energy distributions (SEDs) of stellar populations in distant galaxies \citep[e.g.][]{2022MNRAS.512.5329B,2022ARA&A..60..455E,2023ApJ...958..141L, 2024MNRAS.527L...7F}.

However questions and uncertainties remain. In particular, recent work has highlighted
the fact that there is excess nebular \heii\ emission seen in rest-frame ultraviolet spectra of high-redshift galaxies \citep{Senchyna2020, Berg2019}. This line has an ionization potential of 54.4\,eV, and thus requires an extremely blue (hot) ionizing spectrum from the powering source. As we have pointed out before \citep{Stanway2016}, the contribution of XRBs to this line flux may be important, particularly since initial mass function (IMF), variations are unlikely to produce the observed emission \citep{2019A&A...621A.105S}. While some research \citep[][]{Schaerer2019} suggests that the contribution of XRBs to this line flux is significant, (showing that the  \heii\ $\lambda4686$ intensity and trend with metallicity can be reproduced by the He ionizing photons from X-ray binaries), and that such emission can extend over significant areas \citep{Soria}, others have suggested it is unlikely to be able to explain all the line flux or that including a substantial XRB population would result in an overestimate of the associated emission in hydrogen recombination lines \citep[e.g.][]{Senchyna2020, Kehrig2015}. {\cite{Senchyna2020}, using the multi-color disc model of \cite{Mitsuda1984} and calculating the resulting spectra using the \textsc{ARES} package of \cite{Mirocha2014} contend that the predicted \heii\ / He$\beta$ line ratios vs X-ray luminosity cannot be reproduced by photoionization models. (We discus this in more detail in Section \ref{nebemislines}). Such debates highlight the importance of calculating an XRB emission model that is, by construction, motivated by and constrained by the stellar population that dominates other aspects of the nebular emission spectrum (including the hydrogen lines).

While past releases of \textsc{BPASS} stellar populations have included a model of accretion onto compact objects forming XRBs \citep[see][for details]{Eldridgeetal2017} the electromagnetic (EM) emission associated with accretion in these systems has not been included in the spectral synthesis. 

The \textsc{BPASS} framework (comprising detailed binary stellar evolution models, their population synthesis and a spectral synthesis), retains key information about the lifetime of individual stellar binary models, their mass transfer histories, compact remnant masses and continued evolution after the formation of a remnant. It is thus extremely well suited to self-consistently modelling the XRB accretion SED, simultaneously with the same population's stellar SED. To this end, in this paper we describe v2.2.\textsc{X-BPASS}, a planned supplementary release of \textsc{BPASS} that calculates the emission from X-ray binaries, allowing this to be combined with the synthetic stellar spectra of the same stellar population from which the XRBs arose. 

The outline of this paper is as follows: in Section \ref{sec:nummethod} we give a very brief outline of the \textsc{BPASS} modelling framework relevant to this research, describe our treatment of the different types of accretion in X-ray binaries, how we calculate the temperature and luminosity of each X-ray binary system and how these are combined to estimate a stellar population's X-ray flux. In Section \ref{sec:modval} we investigate the number of XRBs predicted for a given stellar population and compare this to observational constraints from the Small Magellanic Cloud (SMC) \citep{Antoniou2019}. We also calculate the X-ray flux evolution with time  and compare our results with observational constraints in the galaxy M51  \citep{Lehmer2017}, investigating the effects of metallicity on these distributions. 
In Section \ref{sec:specsynth}, we discuss the  impact of X-ray emission on the environment of stellar populations, and take a closer look at the expected number of photons generated above the H\,{\sc I}, \heii\ and O\,{\sc VI} ionization thresholds by a stellar population, in light of recent constraints from observations of high red-shift galaxies. Finally, in Section \ref{discuss} we highlight caveats to this work, review our findings and outline future work before presenting a summary of our conclusions in Section \ref{conclude}. 

\section{Numerical Method}\label{sec:nummethod}

\subsection{\textsc{BPASS} stellar evolution and population synthesis}

The \textsc{BPASS} framework is built on a large grid of detailed single and binary stellar evolution models. These are calculated at 13 total metallicity mass fractions in the range $Z\ =10^{-5}$ to $Z\ =\ 0.04$ (0.05 - 2\, Z$_\odot$), for primary and single star masses in the range 0.1 - 300\,M$_\odot$, and for binary mass ratio and period distributions informed by local observations. Each stellar model is followed to the end of its evolution, and the mass of any compact stellar remnant (white dwarf (WD), neutron star (NS) or black hole (BH)) is calculated using the internal stellar structure at the end of the model. Similarly, the evolution of any surviving companion, and of the binary parameters (orbital separation, period etc) are also calculated. Mass transfer between binary components is tracked throughout the model evolution (both before and after compact remnant formation). A distribution of possible supernova kicks is sampled and compared to the gravitational binding energy of the binary, to weight the survival and occurrence probability of models in which such events occur. \textsc{BPASS} stellar models use an adaptive time grid, and hence the evolutionary time-steps of individual models vary in duration. For the purposes of population synthesis, results are binned onto a time grid predicting the emission and population properties for ages of $\log_{10}$(age/years) = 6.0 to 11.0, at intervals of 0.1\,dex. 

In this work, we build on the population and spectral synthesis from \textsc{BPASS} v2.2.1 \citep{Eldridgeetal2017,StanwayEldridge2018}. We note the most recent \textsc{BPASS} version, v2.3.1, explores changes in the assumed stellar emission templates but has not altered the underlying v2.2.1 stellar models or population synthesis algorithm. 
Here and throughout the paper, we use the fiducial \cite{Kroupa1993} initial mass function (IMF), which consists of a broken power law with a lower slope of $\alpha=-1.30$ below 0.5\,M$_\odot$ and an upper slope of $-2.35$, extending to a maximum stellar mass of 300 \textup{M}$_\odot$. Unless otherwise stated we use all \textsc{BPASS} metallicities down to $Z=0.00001$. In some plots we exclude the $Z=0.00001$ metallicity. At such a low metallicity the stars tend to be much smaller in diameter and have only weak winds, as a result there are fewer interactions and hence fewer XRB systems with the overall X-ray flux greatly reduced. To include these lower outputs scaling of the plots would be required and comparisons between most metallicities would be unclear. 

For the binary populations, \textsc{BPASS} uses the mass ratio distributions from \cite{M&S2017}. The binary evolution scheme is explained in significant depth in \cite{Eldridgeetal2017}. It was adapted from the scheme in \cite{Hurley2000, Hurley2002}. The main differences to that scheme include that the tides are assumed to only be in effect during the RLOF phase, only one star is evolved in detail at a time with the other star approximated using the equations of \cite{Hurley2000} or as a compact remnant, for stable mass transfer onto a companion star the accretion rate is limited by the thermal timescale of the companion i.e. $M_2/T_{\rm{KN}}$, and common envelope evolution (CEE) is implemented using a gamma-prescription conserving angular momentum rather than an alpha prescription for envelope ejection based on the orbital energy. This results in CEE which is more efficient at unbinding the envelope, leading to less orbital shrinkage and more surviving binary systems \cite[see discussion in][]{Stevance2023}. 

The \textsc{BPASS} models used in this work are freely available from \texttt{https://\textsc{BPASS}.auckland.ac.nz} and \texttt{https://warwick.ac.uk/\textsc{BPASS}}.

\subsection{Accretion treatment overview}\label{bpout}

Within the \textsc{BPASS} framework, an XRB is taken to be any compact remnant (CR), in a binary that is accreting material from its companion. For the purposes of XRB counts we set the detection threshold at an inferred X-ray luminosity above $3.8\times10^{32}$ erg s$^{-1}$ \citep{Antoniou2019}, which is similar to the limit of $10^{32}$ erg s$^{-1}$ used in \cite{Dray_2006}.

A key detail of the current \textsc{BPASS} models is that accretion onto WDs and NSs is limited by the Eddington accretion rate. 
Accretion onto BHs is not limited.

\subsection{Post-processing overview}

Luminosities for classical wind-fed and Roche-lobe overflow (RLOF) systems are calculated using standard wind-fed and RLOF mass accretion rates (Section \ref{sec:modx}), while for Be systems the mass accretion rate and resulting luminosity is calculated following a slightly modified application of the calculations in \cite{Liu2023}. These are outlined in Section \ref{beacc}. For all accretion methods and CR types we calculate the spectrum assuming an accretion disc is formed with the structure outlined in Section \ref{ads} and assuming black-body emission. While not all companions accreting material from Be stars are expected to form accretion discs, observations show that accretion discs are formed in a number of Be NS XRB systems \citep{rast2025} and this method enables a spectra to be created for Be XRBs and added to our total X-ray population. 

Luminosities from over-Eddington mass accretion onto BHs are calculated using the slim disc model of \cite{King2023} explained in Section \ref{sec:spectra}, the implications of which are discussed in \cite{Briel2023} and \cite{vanSon2020}. Although recent work modelling distorted super-critical discs gives arguably more accurate disc temperatures at the inner photosphere radius, spherization radius and outer photosphere radius \citep{Poutanen2007,Garofali2023, Kovlakas2022, Kovlakas2025} the standard slim disc model means a complete disc temperature profile by radius is possible enabling a more comprehensive spectra to be created. 

To calculate the SED we split our accretion disc into 100 annuli, calculate the temperature in the centre of each annuli then use these temperatures to calculate the surface brightness per unit wavelength (see Section \ref{sec:spectra}). We sum the total surface brightness and then scale the total energy output from the surface brightness calculation to match the accretion disc or Be luminosity calculation which, unlike the surface brightness, includes the Eddington luminosity limits, the duty cycle and the radiative efficiency.

Throughout this paper we define the compact remnant as a WD if the mass $M_{\textrm{CR}}$ < 1.4M$_\odot$, as a NS if 1.4M$_\odot$ $\leqslant M_{\textrm{CR}} \leqslant$ 3M$_\odot$ and as a BH if $M_{\textrm{CR}}$ > 3.0M$_\odot$. Accreting WDs exceeding 1.4\textup{M}$_\odot$ are assumed to experience type Ia supernovae and subsequent time steps for these models are removed from our analysis. Accreting NSs exceeding 3\textup{M}$_\odot$ are assumed to experience direct collapse to form BHs. In such circumstances we assume the orbital parameters are unchanged and subsequent time steps are analysed assuming the mass-gaining star is now an accreting BH. \textsc{BPASS} orbits are assumed to be circular and in the case of RLOF, the evolution of the XRB system shifts rapidly towards low or zero eccentricity orbits, as the interactions tend to circularise the orbits. For wind-fed systems (excluding those with Be donors), X-ray emission will most likely only reach the detection threshold limit when the systems are in short orbital periods when circularisation due to tidal interaction is likely to have already occurred. For Be XRBs, eccentricity is a key component in the X-ray emission magnitude. In these cases we assign an eccentricity to our identified Be systems by randomly selecting an eccentricity from the observed Be XRB distribution (see Section \ref{beacc}).

Our stellar evolution models calculate the donor star mass loss rate from both stellar wind and/or RLOF, for each \textsc{BPASS} time step. However, we cannot assume that the resulting CR accretion rate is uniform throughout the time step. Observations show that X-ray binary systems spend a significant amount of time in quiescence, with only short periods in outburst \citep{Advan2023,Tauris2006}. The proportion of time spent in outburst is often referred to as the `duty cycle'.

The percentage of time a typical compact remnant spends in outburst is uncertain and will likely depend on the individual system properties. \cite{Sidoli2018} find an average duty cycle of $\sim 30$ percent for super-giant HMXBs, $\sim 55$ percent for giant HMXBs and $\sim 10$ percent for Be/XRBs.

The physics of the duty cycle is poorly understood and there is no reliable method to assign a duty cycle from the modelled binary parameters. As a first-order approximation, we assume a 20 percent duty cycle and implement this in wind-fed and RLOF systems by reducing the duration of each accretion time step by 80 percent. While somewhat simplistic, we believe this blanket duty cycle provides a reasonable population estimate recognizing that this average value, incorporates the inherent statistical variability where the duty cycle may be significantly more or less than 20 percent as highlighted by \cite{Sidoli2018}. This reduced accretion time-step is then randomly located within the original time-step for the purposes of time binning. 

For Be XRB systems, we follow the prescription of \cite{Liu2023} but individualise the mass accretion rate based on a proportion of the wind-loss of the companion (see section \ref{beacc}). We then calculate the peak luminosity and multiply this by 0.2 to recognise the time spent in outburst.
We apply these methods to all accretion time steps and all CR types, independent of donor or CR mass. Where the resulting accretion episode in a stellar model spans multiple output population synthesis time bins, the relative contributions to each time bin are calculated based on the duration the XRB spends in each time bin. 

\subsection{Mass accretion rates and luminosities}
\label{sec:modx} 

Our procedure for calculating the mass accretion rates and resulting luminosities in a population and spectral synthesis is as follows:

At each step in the \textsc{BPASS} metallicity grid, we search through all secondary models (i.e. those in which at least one star has reached the end of its core-burning lifetime). 
We identify all CRs orbiting their original secondary star and find those main sequence (MS) donor stars that are losing material either by stellar wind or via RLOF. For RLOF and classical wind-fed systems, we assume a steady-state accretion disc is formed around the compact object and that the material accreted in each time step flows through the disc and onto the compact remnant. We assume the light emitted from the compact object originates primarily from this accretion disc and ignore non-thermal sources. Emission in the X-ray is dominated by thermal emission, indeed in some pulsars non-thermal emission is not even detected \citep{Chang}. For these classical wind-fed and RLOF systems we calculate the total disc luminosity for each \textsc{BPASS} time-step from the accretion rate for that time-step using;
\begin{equation}
    L_{\textrm{disc}}= \frac{\textrm{G}M_{\textrm{CR}}\dot{M}_{\textrm{CR}}}{2 R_{\textrm{inner}}},
	\label{eq:ldisc}
\end{equation}

where $M_{\textrm{CR}}$ is the mass of the CR, $\dot{M}_{\textrm{CR}}$ is the accretion rate of the CR assuming the 20 percent duty cycle and $R_{\textrm{inner}}$ is the inner radius of the accretion disc (see Section \ref{ads} for detail on the calculation of $R_{\textrm{inner}}$ and Table \ref{tab:modelparams} for the values used).

Photon production is not the only method the accreting material can lose energy \cite[see][]{LEE2014} and the radiative efficiency of the accreted matter in the electromagnetic spectrum is generally thought to be low, in the range of 10-20 percent \citep{Liu2023}. 
The luminosity is therefore reduced by multiplying by the luminosity by the  radiative efficiency, which we assume is $R_{\textrm{eff}}=0.1$ for WDs and NSs and calculate using the prescription of \cite{Podsiadlowski_2003} for BHs:

\begin{equation}
    R_{\textrm{eff}}= 1- \sqrt { 1 - \left( \frac{M_{\rm CR}(t+\Delta t)} {3 M_{\rm CR} (t)} \right)^2 },
    \label{eq:eta}
\end{equation}

where $t$ and $t+\Delta t$ are the ages at the beginning and end of the \textsc{BPASS} stellar model accretion time-step. 

For all accretion types we calculate the Eddington luminosity and Eddington mass accretion rate using the surface hydrogen fraction of the donor star. For this calculation we again follow the prescription of \cite{Podsiadlowski_2003};
\begin{equation}
    L_{\textrm{Edd}}=\frac{4\pi  \textrm{Gc}  M_{\textrm{CR}}}{0.2 (1 + X)},
	\label{eq:ledd1}
\end{equation}

\begin{equation}
    \dot{M}_{\textrm{Edd}}= 2.6 \times 10^{-7} \textrm{M}_\odot \textrm{yr}^{-1}\left(\frac{M_{\textrm{CR}}}{10\textrm{M}_{\odot}}\right) \left(\frac{0.1}{R_{\textrm{eff}}} \right) \left(\frac{1.7}{1 + X}\right),
	\label{eq:medd1}
\end{equation}

where $X$ is the surface hydrogen fraction of the donor star at the time of mass transfer. 

\subsubsection{Be XRB treatment}\label{beacc}
We identify CR binary systems that are not experiencing RLOF but are losing mass due to stellar winds. These binaries result in either Be or classical wind-fed XRB systems.

We define a Be XRB as a system with a MS donor star whose temperature is greater than 10kK but does not exceed 30kK, has a surface Hydrogen fraction greater than 0.4, has a NS or BH companion (i.e. a CR companion $> 1.39\textrm{M}_{\odot}$), has no He core and has a mass $> 6\textrm{M}_{\odot}$. 

The rotation rate and separation of the Be donor from its companion is also important as these are critical for the creation and stability of the decretion disc. Since \textsc{BPASS} currently does not calculate the stellar rotation rate, we identify the rapid rotation required for a Be star by selecting those XRB systems where the donor has no He core and has experienced a degree of internal mixing indicating rapid rotation. Following \cite{Liu2023}, we require the system to have an orbital period greater than seven days to ensure the viscious decretion disc is not disrupted by tidal forces from the companion but whose period does not exceed 1000 days. In addition, the original mass of the Be star must be $< 30\textrm{M}_{\odot}$ and to account for the spin-down of the Be star, the total mass lost from the Be donor must not exceed 1 percent of the original Be mass. 

For Be XRB systems we use a modified version of the accretion rate formulae of \cite{Liu2023} shown in Equation \ref{eq:mbe}, where we remove their set accretion rate of $4.4 \times 10^{-10} \textrm{M}_{\odot} \textrm{yr}^{-1}$ and replace it with a unique mass accretion rate for each system equal to the wind-loss rate of the donor reduced by a factor of $1 \times 10^{-5}$. The reduction factor is selected to broadly reproduce the observed number of Be XRBs in the Small Magellanic Cloud as outlined in \cite{Antoniou2019}.

\begin{equation}
    \dot{M}_{\textrm{CR}}= \dot{M}_{\textrm{wind}} \times 1 \times 10^{-5} \left(\frac{(1-e)a}{100\textrm{R}_{\odot}}\right)^{-2} \left(\frac{\Sigma_0}{0.015} \right) \left(\frac{M_\textrm{CR}}{1.4}\right)^2,
	\label{eq:mbe}
\end{equation}

Where $\dot{M}_{\textrm{wind}}$ is the wind mass-loss rate from the donor star, $e$ is the system eccentricity, $a$ is the orbital separation and $\Sigma_0$ is the viscosity parameter of the viscous decretion disc given by;
\begin{equation}
    \log(\Sigma_0) = 1.44 \log\left(\frac{M_{\textrm{Donor}}}{\textrm{M}_{\odot}}\right) - 2.37 + \eta(0.0.52),
   	\label{eq:vdd}
\end{equation}

Where $\eta$ (0, 0.52) is a normal distribution with zero mean and standard deviation of 0.52.

Since \textsc{BPASS} assumes circular orbits, we assign eccentricities randomly to our identified Be XRBs based on the best-fit normal distribution to the Be XRBs listed in Table 2 of \cite{Rocha2024}. We experimented with other distributions such as log-normal but found very little difference in the results. 

We then calculate the peak Be XRB luminosity  following \cite{Liu2023};

\begin{equation}
    L_{\textrm{Be}}= \frac{R_{\textrm{eff}}\dot{M}_{\textrm{CR}} c^2}{0.63},
	\label{eq:llui}
\end{equation}

where $R_{\textrm{eff}}$ is the radiative efficiency of the accretor and 0.63 is normalization factor. As per \cite{Liu2023} the peak Be X-ray luminosity is then multiplied by 0.2 to get the timestep averaged luminosity.

\subsubsection{Classical wind-fed XRB treatment}\label{windacc}

We assume all other wind-fed systems that do not fit the Be definition are classical wind-fed systems. For these we assume no minimum mass for the CR companion and assume they accrete material ejected by the wind from the donor star as outlined in \cite{Frank2002} which gives an accretion rate of: 

\begin{equation}
    \dot{M}_{\textrm{CR}}= \frac{1}{4} \times \left(\frac{\dot{M}_{\textrm{wind}}}{0.2}\right) \left(\frac{{M}_{\textrm{CR}}}{M_{\textrm{Donor}}}\right)^2 \left(\frac{R_{\textrm{Donor}}}{a} \right)^2,
	\label{eq:mwind}
\end{equation}

Where the additional 0.2 factor represents the duty cycle. The wind-fed system luminosity is then calculated using Equation \ref{eq:ldisc}

\subsubsection{RLOF XRB treatment}
For CR binaries experiencing RLOF, we assume that the CR mass accretion rate is simply the mass loss rate from RLOF from the donor star modified by the accretion and radiative efficiencies. The resulting disc luminosity is then calculated from Equation \ref{eq:ldisc}. We identify ultra-luminous XRBs (ULXs) as those systems whose luminosity exceeds $10^{39}$ ergs s$^{-1}$ but otherwise these are treated the same as any other super-Eddington accretion system. 

\subsection{Accretion disc structure}\label{ads}
For each CR binary, including Be XRB systems, we use an accretion disc model to calculate the XRB spectrum from the disc inner radius ($R_{\textrm{inner}}$) and outer radius ($R_{\textrm{outer}}$). The inner radius is a free parameter which encompasses much of the uncertainty in our disc model.

It is thought that accretion flows rarely reach down to the stellar surface for massive compact remnants \cite[e.g.][]{King2016, Chiang2016, vandenEijnden2017, Mahmoud2019, Dzielak2019}, although for WDs an $R_{\textrm{inner}} = 1.05 \ \times$ $R_{\textrm{wd}}$ (the WD radius) is not unexpected, \citep{Sion2009}. In this work we select physically-motivated $R_{\textrm{inner}}$ values for each compact remnant type, as described below.

For WDs we set $R_{\textrm{inner}}$, as a multiple of the WD radius which is calculated from \cite{Nauenberg1972};

\begin{equation}
    R_\textrm{wd}= 7.8\times 10^6 \mathrm{m}\left[\left(\frac{1.44}{M_{\textrm{wd}}}\right)^{2/3}-\left(\frac{M_{\textrm{wd}}}{1.44}\right)^{2/3}\right]^{1/2}. 
	\label{eq:wdr}
\end{equation}

For NSs and BHs we determine $R_{\textrm{inner}}$ from \cite{Schwarzschild1916} such that,

\begin{equation}
    R_\textrm{inner}=\frac{\tau  2 G M_{\textrm{CR}}}{\textrm{c}^2},
	\label{eq:rinner}
\end{equation}

where $M_{\textrm{CR}}$ is the mass of the accreting NS or BH and $\tau$ is the number of Schwarzschild radii ($R_{\textrm{s}}$) at which the inner disc is truncated. Smaller $\tau$ values lead to more X-ray emission and larger values lead to less. 

We set the default inner truncation radius as follows (see Table \ref{tab:modelparams}); for WDs we use $R_{\textrm{inner}}$ = 1 WD radius, for NSs we use $\tau = 8$, and for BHs we use $\tau = 3$. Our inner truncation radius for NSs is in good agreement with the values \cite{King2016} and \cite{Ludlam2017} find for the NS accretor in Aquila X-1 of 7.5$\pm$ 1.5 $R_\textrm{s}$ and 5.5$^{+1}_{-0.5}$ $R_\textrm{s}$ respectively. $R_{\textrm{inner}}$ will also be dependent on the magnetic field which will vary between individual NSs and WDs. \textsc{BPASS} does not currently calculate magnetic fields so our inner truncation radius for these objects represents the average population effects of magnetic fields. For BHs our value of $\tau = 3$ means the $R_{\textrm{inner}}$ is set at the value for the inner most stable circular orbit ($R_{\textrm{ISCO}}$). 

Next we define the disc outer radius. For this we assume the lesser of the radius of the L1 and L2/L3 Lagrange points of the CR. While these Lagrangian points are defined for spherically-symmetric accretion we use them as a first order approximation here. We have explored a range of other assumptions but find our results are only weakly dependent on this parameter since the emission is dominated by the hot central regions of the disc. We further impose a minimum outer radius for all compact remnants so that $R_{\textrm{outer}}$ must be at least twice that of $R_{\textrm{inner}}$.

\begin{table}
    \centering
    \begin{tabular}{l|ccc}
        Remnant & BH & NS & WD \\
        \hline\hline
        Mass / M$_\odot$ & $> 3.0$ & 1.4 -- 3.0 & $< 1.4$\\
        Accretion Limit & $>\,L_{\textrm{Edd}}$ (Eqn \ref{eq:lslim1}) & $L_{\textrm{Edd}}$ & $L_{\textrm{Edd}}$\\
        $R_\textrm{inner}$ & 3\,$R_s$ & 8\,$R_s$ & 1\,$R_\mathrm{wd}$\\
        $R_\textrm{outer}$ & min\{L1:L2\slash L3\} & min\{L1:L2\slash L3\} & min\{L1:L2\slash L3\}\\
        $R_{\textrm{efficiency}}$ & See Eqn \ref{eq:eta} & 0.1 & 0.1 \\
        Duty Cycle & 0.2 & 0.2 & 0.2 \\
    \end{tabular}
    \caption{Summary of model parameters for different compact remnant types.}
    \label{tab:modelparams}
\end{table}

The next step is to calculate the temperature profile of the accretion disc. To do this we split the disc into 100 annuli including one at the maximum temperature position of the disc at 49/36 $R_{\textrm{inner}}$. For all CRs exceeding the Eddington luminosity, the temperature of each annulus is calculated using the Eddington mass accretion rate. For CRs accreting below the Eddington luminosity the annulus temperatures are calculated using the actual mass accretion rates. 

 We calculate the temperature in the centre of each annulus using the formula of \cite{Padmanabhan2001};

\begin{equation}
    T(R)=\left[ \frac{3\textrm{G}M_{\textrm{CR}}\dot{M}_{\textrm{CR}}}{8\pi R^3 \sigma}\left(1-\frac{R_\textrm{inner}}{R}\right)^{\frac{1}{2}}\right]^{\frac{1}{4}}.
	\label{eq:tanul}
\end{equation}

Where the annulus temperature exceeds 10kK (approximately the hydrogen dissociation temperature), we introduce the spectral hardening factor calculated by \cite{Shimura1995}. For temperatures in excess of 11kK we multiply the calculated temperature by 1.7. For temperatures between 10kK and 11kK, to ensure a smooth transition, we increase the factor from 1.0 at 10kK to 1.7 at 11kK using the cosine function.

We considered whether the proximity of the accretion disc to the BH would induce a significant red-shift in the emitted radiation, thus reducing the energy budget for the accretion disc. For the inner truncation radius adopted for the BHs ($R_{\textrm{inner}}$ = $R_{\textrm{ISCO}}$) the red-shift is only of the order of 6 percent for the hottest part of the disc, and hence the overall effect would be insignificant. We also ignore the effect of the X-ray heating of the disc from the compact remnant \citep{vanPara1994,lasota2023}. Due to the very small radii of NSs and BHs, this is most likely only to be of significance for WDs. However, since very few WDs reach the XRB luminosity threshold, this effect will have a negligible effect on our analysis. We also currently neglect  the effects on the evolution of the companion star as a result of irradiation from the CR \citep{Tout1989}, but aim to investigate this effect further in future work.

\subsection{Luminous Output}
\label{sec:spectra} 
We use the temperatures of the annuli  calculated in Equation \ref{eq:tanul} in Section \ref{ads} to calculate the surface brightness per unit wavelength assuming a blackbody spectrum using;
\begin{equation}
    B_{\lambda}(T)= \frac{2\textrm{hc}^2}{\lambda^5} \left[\textrm{exp} \left(\frac{\textrm{hc}}{\lambda k_BT}\right)-1\right]^{-1}
	\label{eq:sb}
\end{equation}
where $k_B$ is the Boltzmann constant. 

We consider wavelengths from 0.1 to 1000 \AA ngstroms in increments of 0.1\AA\  and from 1100 to 10000 \AA ngstroms in increments of 100\AA\ . For each annulus, we calculate the luminosity in each wavelength bin by multiplying the surface brightness by the bin width and annulus area and then sum the bins to obtain the total disc luminosity.

For WD and NS models, the total surface brightness calculated is scaled to match the lesser of the disc luminosity calculated in Equation \ref{eq:ldisc} or the Eddington luminosity calculated in Equation \ref{eq:ledd1}. For BHs emitting below the Eddington luminosity the surface brightness is also scaled to match the disc luminosity calculated in Equation \ref{eq:ldisc}. For BHs exceeding the Eddington luminosity, the luminosity in each \AA ngstrom bin is scaled to ensure the total disc luminosity from the surface brightness calculation matches the slim disc model of \cite{King2023} calculated in Equation \ref{eq:lslim1}. 

\begin{equation}
    L_{\textrm{disc}} = L_{\textrm{Edd}}\left[1 + \textrm{ln}\left(\frac{\dot{M}_{\textrm{CR}}}{ {\dot{M}_{\textrm{Edd}}}}\right)\right]
	\label{eq:lslim1}
\end{equation}

Finally, since the \textsc{BPASS} time bins are not of uniform duration, we bin the XRB numbers and flux into time bins evenly distributed in log space from 6Myrs to 13Gyrs. We reiterate that for Wind-fed and RLOF systems for the purposes of the time binning, we assume one shortened accretion episode based on the duty cycle, and randomly locate this in the original time-step. If this reduced time-step spans more than one of the new time bins, the total flux from the accretion episode is distributed proportionately over each time bin. For Be XRBs the reduced peak fluxes are assumed across the entire original time step and then proportionately distributed over the new time bins.

\section{Model validation}\label{sec:modval}
\subsection{Small Magellanic cloud (SMC)}
\label{sec:m51} 

We validate our model construction through comparison with observational data. \cite{Antoniou2019} compiled a census of HMXBs, in the Small Magellanic Cloud. This low metallicity dwarf satellite of the Milky Way is experiencing a galaxy-wide starburst, and includes sites of high-mass star formation. \cite{Antoniou2019} used their data to derive a HMXB rate per unit stellar mass, and per unit star formation rate, as a function of time, assuming a metallicity of $Z=0.004$. We caution that the stellar population masses and ages used by \cite{Antoniou2019} did not take into account interacting binaries (i.e. were obtained from fitting a single star SED model to the stellar population), and hence the observationally-inferred uncertainties shown on reported data may be underestimated. We use these data nonetheless to comment on trends in our data and the approximate XRB population number.

In Figure \ref{fig:XRBnumbers2_8}, we compare the results derived by \cite{Antoniou2019} to the predicted model number of HMXBs per stellar age time bin, for 10$^6$ M$_{\odot}$ of star formation at the \textsc{BPASS} metallicity of $Z=0.004$. Figures \ref{fig:XRBnumbers004} and \ref{fig:XRBnumbersv2} show the predicted total number of XRBs split by CR type and donor mass respectively for the \textsc{BPASS} metallicity of $Z = 0.004$. In each figure, the results from \cite{Antoniou2019} are shown as black symbols with error bars. The figures show that the \textsc{BPASS} predictions at $Z = 0.004$ are in generally good agreement with the HMXB observations in the SMC although we overpredict the XRB numbers at ages less than 7.5 log(age/years). 

\begin{figure}
\centering
    \includegraphics[width=\columnwidth]{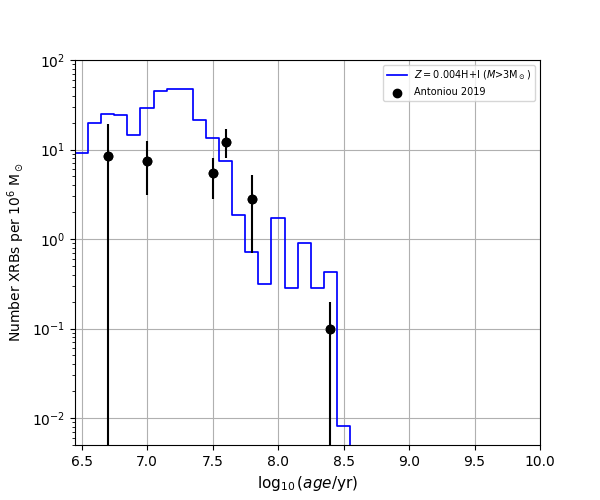}
    \caption{HMXBs expected in each time bin for $10^6$ M$_\odot$ of stars formed from the fiducial \textsc{BPASS} IMF metallicity $Z = 0.004$. For HMXBs we have assumed a mass greater than 3M$_{\odot}$. The black circles represent the number of HMXBs and the black lines represent the uncertainties derived by \citet{Antoniou2019} from various fields in the SMC for which they assumed a metallicity of $Z = 0.004$. The point in the log(age/years) = 8.3 time bin is an upper limit on the number of HMXBs. In our model predictions, we include all BH and NS X-ray binaries that have an X-ray luminosity of $3.8\times10^{32}$\,ergs\,s$^{-1}$ or greater; this is the limiting flux achieved by \citet{Antoniou2019}.}
    \label{fig:XRBnumbers2_8}
\end{figure}

The distinction between HMXBs and low-mass XRBs (LMXBs), is based on the mass of the donor star. Observationally, HMXBs cannot be easily distinguished from LMXBs \citep{Lehmer2017, Mineo2011}, and there is no universally accepted donor mass threshold between the two categories. For example \cite{Tauris2006} define the limit for an HMXB donor star of $M_\textrm{donor}$ > 10M$_\odot$ and a LMXB limit of $M_\textrm{donor} \leqslant$ 1.5M$_\odot$ with intermediate mass XRBs (IMXB) populating the mass range 1.5M$_\odot$ < $M_\textrm{donor} \leqslant$ 10M$_\odot$, while \cite{Hunt2021} define HMXBs as those systems where $M_\textrm{donor} \geqslant$ 8M$_\odot$, an IMXB with 3 $\leqslant$ $M_\textrm{donor}$ < 8M$_\odot$ and LMXB with $M_\textrm{donor}$ < 3M$_\odot$. By contrast, \cite{Lehmer2017} use a mass limit for HMXBs of $M_\textrm{donor} \geqslant$ 2.5M$_\odot$, and other researchers such as \cite{Prestwich2003} use the X-ray spectra to differentiate HMXBs from LMXBs. 

For the purposes of our analysis, we adopt the definitions of \cite{Hunt2021} and in addition we place a luminosity limit defining an XRB as a binary system whose  luminosity from accretion exceeds $3.8\times10^{32}$ erg s$^{-1}$. Since many researchers do not separate out IMXBs we include these in the HMXB numbers for our comparisons with \cite{Antoniou2019}. While we include WD XRBs in this plot, we note that their local column density is typically significantly greater than that of NSs and BHs, so most of the X-rays are trapped within the accretion disc. As a result, WDs are generally only weak X-ray sources \citep{Tauris2006}, and very few reach the XRB threshold of $3.8\times10^{32}$ erg s$^{-1}$. 

The XRB population, split by donor mass in  Figure \ref{fig:XRBnumbersv2} demonstrates that the number of sources excluding those classified as LMXBs, provide a better match to the source numbers inferred by \cite{Antoniou2019} than the total accreting population or the LMXB population alone. However, we highlight that our split between HMXB, IMXB and LMXB numbers \cite[as well as those of][]{Antoniou2019}, are dependent on what mass cutoff points are used.  

Figure \ref{fig:XRBnumbersv2} demonstrates that HMXBs in both observations and models typically have ages less than 100Myrs. The number of IMXBs peak at approximately 30Myr then reduce steadily until 10Gyr.   

In contrast, LMXBs do not begin to appear until approximately 13Myrs but dominate (in numbers if not necessarily luminosity), at ages greater than approximately 300Myrs. 
 
\begin{figure}
\centering    \includegraphics[width=\hsize]{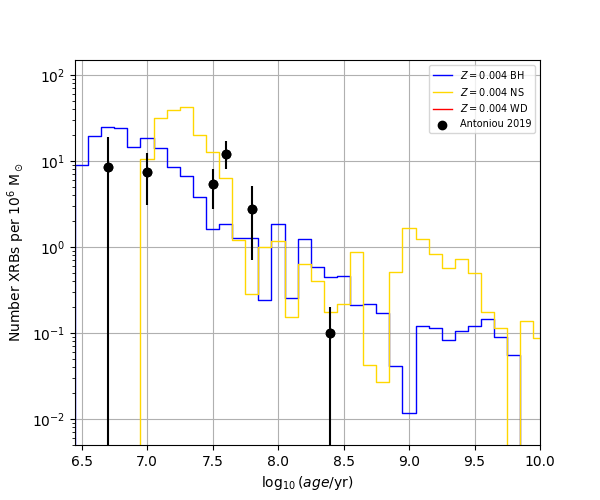}
    \caption{All XRBs expected in each time bin for $10^6$ M$_\odot$ of stars formed from the fiducial \textsc{BPASS} IMF for a metallicity of $Z = 0.004$ split by compact remnant type.}
    \label{fig:XRBnumbers004}
\end{figure}

\begin{figure}
\centering
    \includegraphics[width=\hsize]{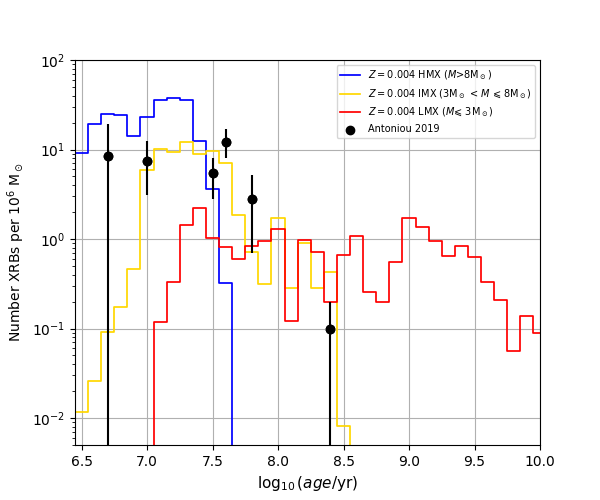}
    \caption{As in Figure \ref{fig:XRBnumbers004}, for metallicity of $Z = 0.004$ split by donor type.} 
    \label{fig:XRBnumbersv2}
\end{figure}

\subsection{Effect of metallicity on XRB numbers}

In Figure {\ref{fig:numsall}} we show the XRB numbers formed from the fuducial \textsc{BPASS} IMF and metallicities split into three metallicity bands. In the top panel we show \textsc{BPASS} metallicities from $Z = 0.03 -0.008$, in the centre $Z = 0.008 -0.002$ and in the bottom panel $Z = 0.002- 0.0001$. We show the numbers from \cite{Antoniou2019} in all panels but note that these were for the SMC, which has an estimated metallicity of $Z = 0.004$. The \cite{Antoniou2019} data is  included to highlight the relative metallicity differences and not as a direct comparison. We caution that for ages greater than $\approx$100Myr there are very few BPASS stellar models contributing to the XRB population. In Figure {\ref{fig:numserr}} we highlight this by showing the Poisson uncertainties associated with Figure {\ref{fig:numsall}}.
Uncertainties are shown as a percentage of the predicted compact remnant numbers, calculated using 1/$\sqrt{N}$ where $N$ is the number of models contributing to the number counts in each time bin. We note that for metallicities greater than $Z = 0.008$, at times prior to 10Myrs and later than 100Myr, the number of accreting binary systems is low and as a result the uncertainties are significant (greater than 100 percent of the predicted number counts in many cases). For metallicities lower than $Z = 0.008$, the number of XRBs at later times is also low and the uncertainties exceed 100 percent at around 100 Myrs.  

\begin{figure}
\centering
   \includegraphics[width=\hsize]{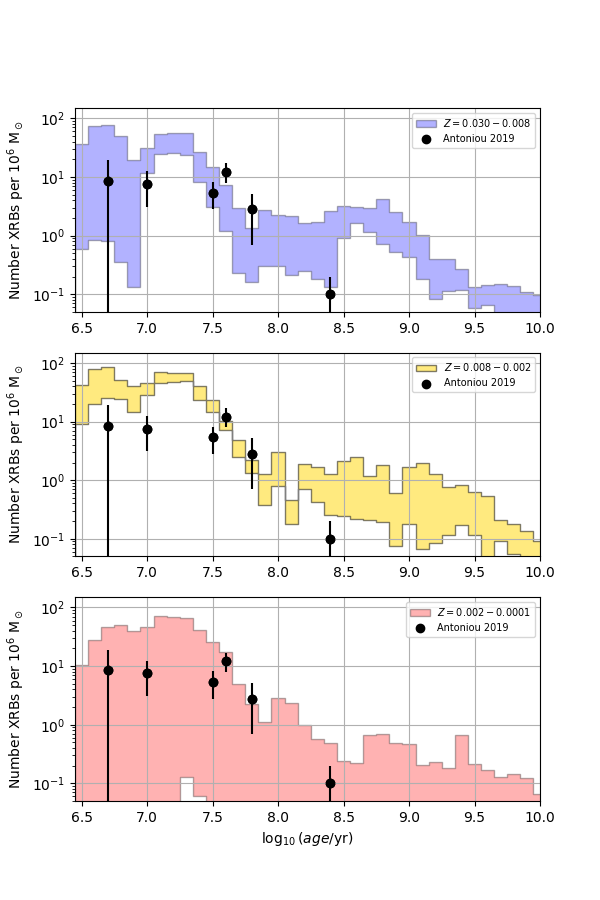}
    \caption{The number range of XRBs expected in each timebin for 10$^6$ M$_{\odot}$ of stars formed from  fiducial \textsc{BPASS} metallicities split into three metallicity bands. Top panel $Z = 0.03$ to $Z = 0.008$, centre panel $Z = 0.008$ to $Z = 0.002$ and bottom panel $Z = 0.002$ to $Z = 0.0001$. Numbers from \citet{Antoniou2019} for the SMC ($Z=0.004$) are shown in each panel to highlight the comparative differences.}
    \label{fig:numsall}
\end{figure}

\begin{figure}
\centering
   \includegraphics[width=\hsize]{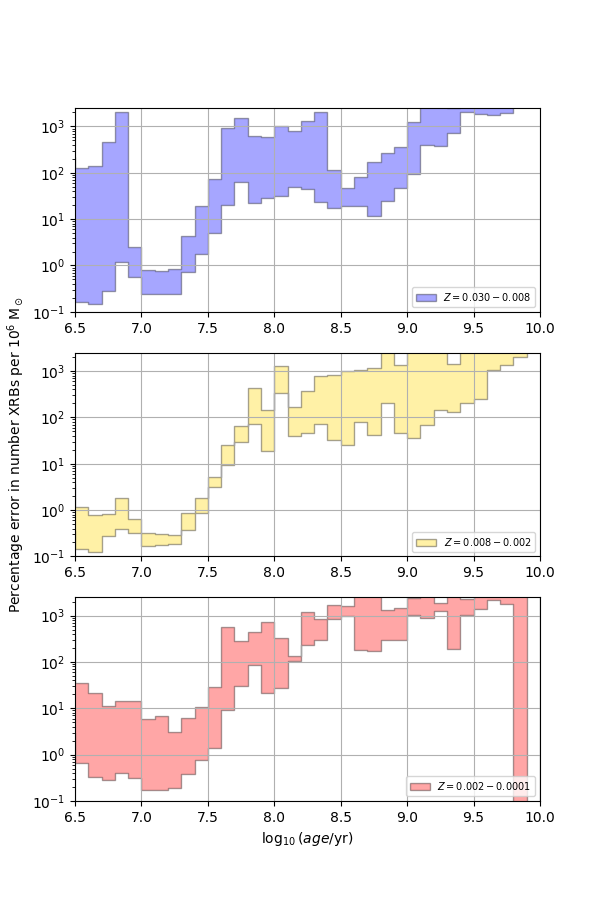}
    \caption{As per Figure \ref{fig:numsall} but now showing the Poisson uncertainties associated with the XRB number ranges shown in Figure \ref{fig:numsall}.}
    \label{fig:numserr}
\end{figure}

The stellar evolution for metallicities below $Z=0.001$ is significantly different and as a result the increasing HMXB trend does not continue below $Z=0.001$. This is evident in the lower panel of Figure {\ref{fig:numsall}}

 We also find that for each $10^6$\,M$_\odot$ of stars formed, the total number of XRBs in a population is of order 1 to 10 in the age range 30\,Myr - 1\,Gyr until metallicities below $Z=0.002$ are reached where these numbers drop below 1 after approximately 300Myr.

We now consider the XRB numbers split by donor mass shown in Figure \ref{fig:numstype}. \cite{Dray_2006} studied simulations of HMXBs as a function of metallicity, suggesting that both the number and the mean period of HMXB progenitors can vary with metallicity, resulting in increased numbers of HMXB systems in sub-Solar metallicity stellar populations.
The variation of HMXB numbers with metallicity suggested by \cite{Dray_2006} is clearly evident in the top two panels of Figure {\ref{fig:numsall}} and in the top panel of Figure {\ref{fig:numstype}}. The predicted number of XRBs in a fiducial $10^6$\,M$_\odot$ population  is similar for all metallicities, with a general pattern of more HMXBs and fewer LMXBs observed as metallicity reduces. We see that LMXBs dominate at late ages and there appears to be significant variation with the initial metallicity of the stars with higher metallicities generally producing more LMXBs (see bottom panel in Figure \ref{fig:numstype}). However, the trend is less clear for the HMXBs and IMXBs where the maximum number in any one time-bin reduces but the average number of HMXBs and IMXBs increases. It would also appear that the number of IMXBs, like HMXBs, reduce below metallicities of $Z=0.001$ (see top and centre panels of Figure \ref{fig:numstype}). 

We have attempted to find recent predictions of the birth rate of LMXBs. \cite{Kalogera1998} explored an analytic population synthesis, informed by a small grid of stellar evolution models, and estimated that the birth rate of LMXBs in our Galaxy is $2\times10^{-7}$ yr$^{-1}$ based on an analysis of then-extant Galactic LMXB catalogues. This was consistent with their models assuming a canonical total star formation rate for the Milky Way of about 2-3\,M$_\odot$\,yr$^{-1}$. 

A more recent study by \cite{Lehmer2020} analysed the luminosity function of XRBs, focussing on those associated with the globular cluster populations of elliptical galaxies. These are old stellar populations with typical ages of 5-10\,Gyr. As such any X-ray emission is deemed to be associated with LMXBs rather than their high-mass counterparts. However, their luminosity threshold is often applied to the entire emission from an unresolved population rather than to individual stars within that population, making a direct comparison to our models challenging. 

\begin{figure}
\centering
   \includegraphics[width=\hsize]{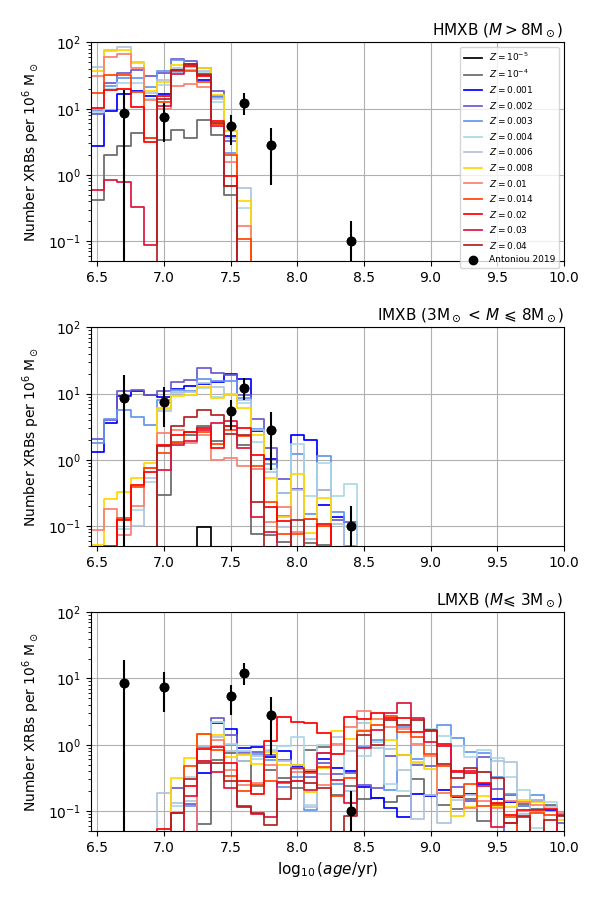}
    \caption{The number of XRBs expected in each timebin for 10$^6$ M$_{\odot}$ of stars formed from the fiducial \textsc{BPASS} metallicities split by donor mass. Top panel HMXBs, middle panel IMXB and bottom panel LMXBs.}
    \label{fig:numstype}
\end{figure}

Another key observable validation point is how the integrated total X-ray luminosity of a stellar population varies with age. We show in Figure \ref{fig:flux0402} our prediction for the X-ray flux from 2 to 7 keV for \textsc{BPASS} metallicities between $Z = 0.04$, and $Z = 0.01$. This is the same energy window as that adopted in an observational study by \cite{Lehmer2017} on the Whirlpool Galaxy, M51, and broadly consistent with the range of estimated metallicities in the galaxy's disc \citep[which shows a strong metallicity gradient, see e.g.][]{2020PASP..132i4101W}. \cite{Lehmer2017} claim that the metallicity of M51 is at least twice solar, using one of the values derived in \cite{Moustakas2010} for the galaxy. However, metallicity indicators are very uncertain especially as binary stars are typically ignored in the analysis \citep[see][for example]{Kewley2008, Xiao2018}. A detailed study of M51 by \cite{Bresolin2004} found a sub-solar metallicity for M51 and one of the values given by \cite{Moustakas2010} is similar. 

The time evolution derived from our model grid shown in Figure \ref{fig:flux0402} at these metallicities is similar to that derived by \cite{Lehmer2017}, showing a gradual decline in flux after 10\,Myr with both the order of magnitude and the long term gradient in source flux per stellar mass consistent with the observational data, given their associated uncertainties. 

Extending the comparison to all \textsc{BPASS} metallicities, Figure \ref{fig:fluxall}, demonstrates that the rate of decline of X-ray flux is not strongly metallicity dependent after log(age/years) $>9$. The higher metallicities (generally) have lower X-ray flux at early times due to higher mass loss giving rise to a reduced number and lower mass, black holes. This also results in wider binaries that interact more weakly and a flatter time evolution of the integrated flux. The rarity of bright accretors at these metallicities leads to more stochasticity in the modelled X-ray flux as relatively few stellar models are dominating the emission. By contrast the lowest metallicity models would over-predict the early time flux and rate of X-ray evolution observed in M51 (as would be expected given its near-Solar composition). Interestingly the X-ray fluxes per solar mass converge at ages beyond 3\,Gyr, in the LMXB-dominated regime. 

\begin{figure}
\centering
    \includegraphics[width=\hsize]{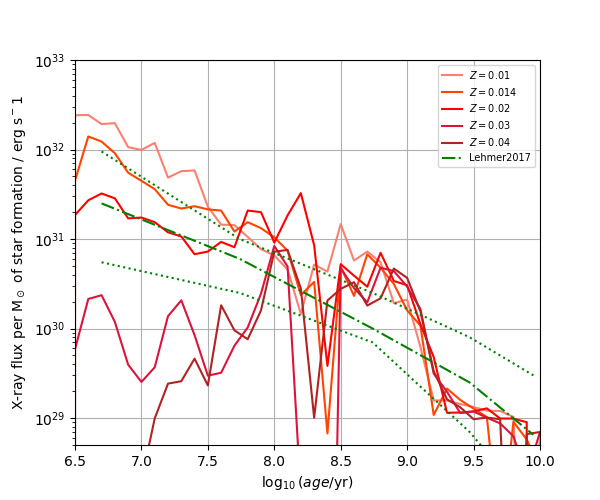}
    \caption{X-ray flux per Solar mass of star formation in ergs s$^{-1}$ for \textsc{BPASS} metallicities $Z = 0.04$ to $Z = 0.01$ compared to the X-ray flux distribution for M51 derived by \citet{Lehmer2017} who suggest a metallicity of M51 of 1.5 - 2.5 $Z_{\odot}$. The green dot-dash line is the best fit with dotted lines for uncertainties.}
    \label{fig:flux0402}
\end{figure}

\begin{figure}
\centering
   \includegraphics[width=\hsize]{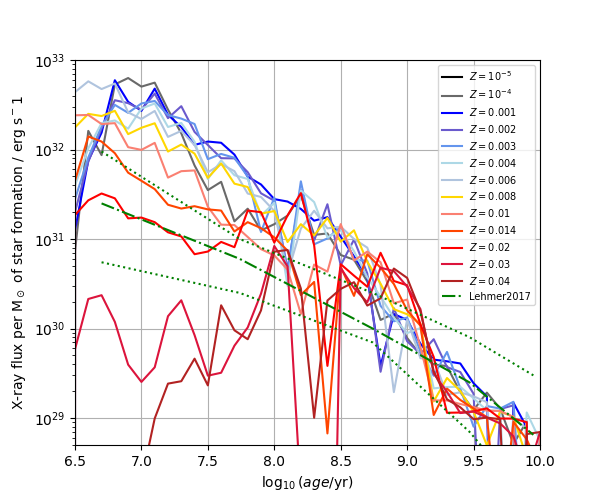}
      \caption{X-ray flux per Solar mass of star formation in ergs s$^{-1}$ for \textsc{BPASS} metallicities, compared to the X-ray flux distribution for M51 derived by \citet{Lehmer2017}. The green dot-dash line is the best fit with dotted lines for uncertainties.}
         \label{fig:fluxall}
\end{figure}

M51 is a grand-design spiral galaxy in the process of an ongoing merger, and hosting a central active galactic nuclei (AGN). However, detailed analysis of any single galaxy can be misleading if its star formation history, metallicity and evolutionary state are not fully understood. As a further validation of our models we compare our X-ray flux  with the best-fit X-ray flux function by age for a larger sample of normal, star-forming galaxies identified in the Chandra Deep Field-South and North surveys and the Great Observatories Origins Deep Survey \citep[GOODS,][]{Giavalisco2004}. These deep survey fields benefit from extensive multi-colour photometry and extremely sensitive wide-area X-ray observations. Thus, while any individual galaxy may be barely resolved and have detections in a limited number of bands, stacking analyses and statistical studies can recover the behaviour of entire galaxy populations.

\cite{Gilbertson2022} used GOODS fields to study both LMXB and HMXB X-ray flux by stellar population age in the 2-10 keV band, for 344 galaxies with spectroscopic redshifts in the range $Z=0-3.5$. Selection biases in the deep fields result in a bias towards star-forming stellar populations. The stellar population ages and masses for each galaxy were determined through galaxy SED fitting, using models built from the PEGASE single star stellar library. The median metallicity of the sample is $0.63Z_{\odot}$ \citep{Garofali2023}. 

In their analysis, \citet{Gilbertson2022} derived a fitting function for the ratio of X-ray flux to stellar mass as a function of stellar population age, which is comparable to the results of \citet{Lehmer2017} for the detailed study of M51. In Figure \ref{fig:GILB} we show the comparison of the \cite{Gilbertson2022} best-fit X-ray flux to stellar mass ratio evolution (with LMXB and HMXB flux combined) to the \textsc{BPASS} models for $Z=0.006$ to $Z=0.02$ which covers the metallicity range of the galaxies used  ($0.4 - 1.16Z_{\odot}$). Other than an excess flux from our models around log(age/years)= 8.5 - 9, we find a good fit with the \textsc{BPASS} models. We do not show the uncertainties estimated by \cite{Gilbertson2022} which were calculated  using models with the lowest 68 percent of the fitting `C' values and were not provided in the paper. They comment that the uncertainties in their best-fit functions are considerably higher if the full range of models used in their fitting `C' values are considered.

\begin{figure}
\centering
   \includegraphics[width=\hsize]{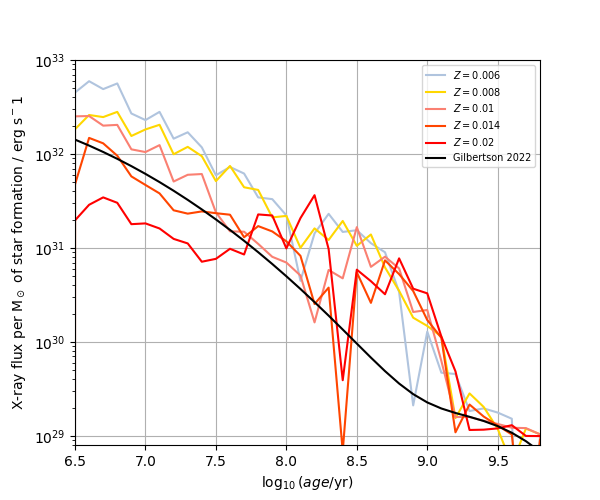}
      \caption{X-ray flux per Solar mass of star formation in ergs s$^{-1}$ for \textsc{BPASS} metallicities $Z=0.006 - 0.02$, compared to the best fit X-ray flux distribution from \citet{Gilbertson2022}. LMXB and HMXB combined flux is shown as a solid black line.}
      \label{fig:GILB}
\end{figure}

\begin{figure}
\centering   \includegraphics[width=\hsize]{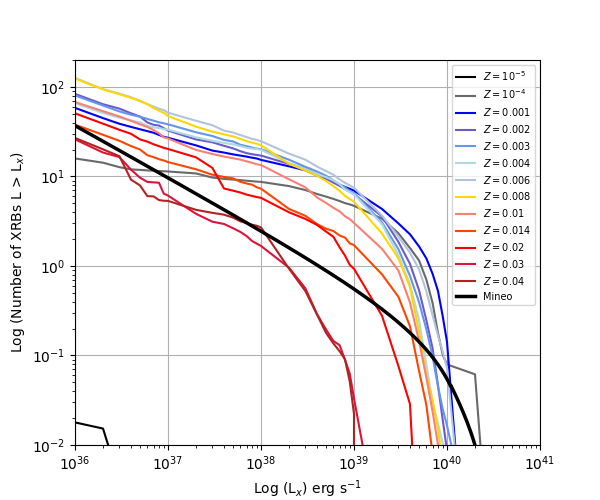}
    \caption{The cumulative X-ray luminosity function  for the 13 fuducial \textsc{BPASS} metallicities per unit star formation rate. The black line shows the best fit broken power law of \citet{Mineo2011} for their primary sample of 29 star forming galaxies within 40Mpc}
    \label{fig:xlf}
\end{figure}

\subsection{X-ray luminosity function}\label{xlumf}

In Figure \ref{fig:xlf} we show the X-ray luminosity functions (XLFs), for the 13 fudicial \textsc{BPASS} metallicities per unit star formation rate (SFR), compared to the best-fit broken power law of \cite{Mineo2011} for their primary sample of 29 star-forming galaxies within 40Mpc. While our models are in broad agreement with their broken power-law, we create more XRBs in the $10^{36}$ to $10^{40}$ erg s$^{-1}$ range but fewer XRBs at energies above $10^{40}$ erg s$^{-1}$. Some of this shortfall is a consequence of the \textsc{BPASS} version we have used limiting the accretion onto NSs to the Eddington limit. To test this, we ran an updated version of our post-processing code that added super-Eddington accretion onto NSs and re-calculated the X-ray luminosities and SEDs (see Appendix \ref{a1}). The result is a significant increase in the HeII / H$\beta$ line ratio as well as an increase in X-ray flux generally. 

Our post-processing approach is also likely to underestimate the additional X-ray flux as accretion onto NSs from a more massive companion would shrink the orbit (which the post-processing code does not re-calculate), resulting in more significant accretion rates and luminosities in subsequent time steps pushing models further into the ULX range (see Appendix \ref{a1}).
 
\begin{figure}
\centering
   \includegraphics[width=\hsize]{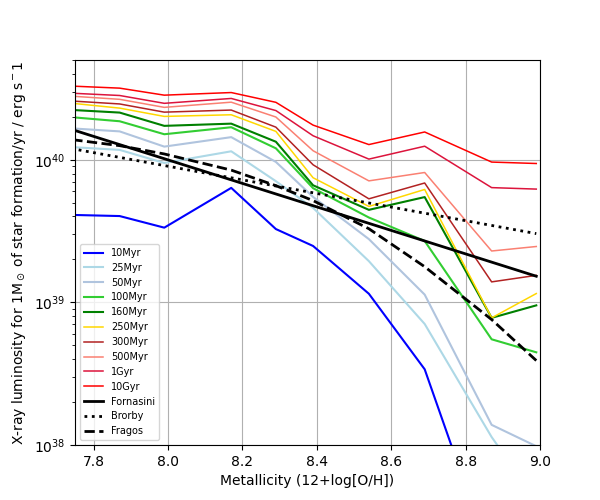}
    \caption{The predicted X-ray luminosity at different metallicities for our \textsc{BPASS} populations from 2-10keV. We show the \textsc{BPASS} luminosity evolution at the population ages shown in the legend and compared with observations}
    \label{fig:xrayevo}
\end{figure}

\begin{figure}
\centering   \includegraphics[width=\hsize]{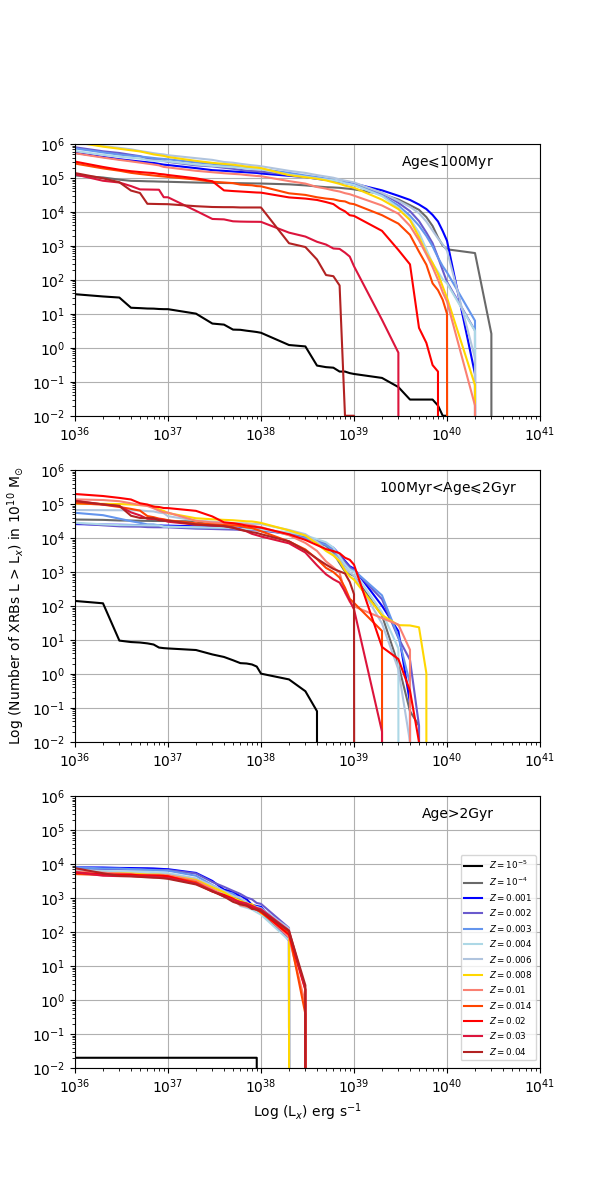}
    \caption{The cumulative X-ray luminosity function for a population of $10^{10}$ M$_{\odot}$ for the 13 fuducial \textsc{BPASS} metallicities split into three time bins.}
    \label{fig:xlf_age}
\end{figure}

\subsection{X-ray luminosity as a function of age and metallicity}\label{mettrend}

While the validation examples above have focused on the evolution of XRB luminosity with stellar population age at fixed (or assumed fixed) metallicity, others have explored the more complex evolution of XRB luminosity with both stellar population age and stellar metallicity simultaneously, either from an observational \citep[e.g.][]{Douna2015, Brorby2016, Ponnada2019,Fornasini2019, Fornasini2020, lehmer2024} or theoretical and modelling 
\citep[e.g.][]{Fragos2013a, Fragos2013b, Douna2015,Brorby2016, Madau2017} perspective. 

The metallicity / X-ray luminosity relationship is inherently difficult to estimate observationally for two reasons. Firstly, as we saw in Section \ref{sec:m51}, the metallicities of stellar populations are difficult to accurately measure \citep[e.g.][]{Kewley2008}. To compare the XRB luminosity per unit of formed mass, the total star-formation rate must also be calculated. However, the variation of star-formation rate indicators which themselves evolve with metallicity, are often not taken into account. \cite{Eldridgeetal2017} and others have shown that there is significant variation in the rest-UV and optical flux with the metallicity of the stellar population, and if this isn’t taken into account in the star formation rate indicator calibration then a systematic error is introduced into any calculations of the star-formation rate.

Hence, while all the studies cited above show a clear trend of increasing X-ray flux with lower initial metallicity, the actual shape of the increase with metallicity is uncertain and there might be more complex structure than observers have suggested to date. 

To explore this question we calculate how the X-ray luminosity of our stellar populations vary with metallicity under the assumption of constant star-formation rate and attempt to compare our predictions to those of \cite{Fragos2013a}; \cite{Brorby2016}; \cite{Madau2017} and \cite{Fornasini2020}. Our results, illustrated in Figure \ref{fig:xrayevo}, show that our models reproduce the same increase in X-ray luminosity with decreasing metallicity seen in previous work. We highlight that while the UV-optical flux in most stellar populations stabilises after an elapsed constant star formation rate interval of 100\,Myr, this is not the case for X-ray luminosity. However, we should highlight that in these  galaxies IMXBs and LMXBs, which are much older, contribute virtually nothing to the measured X-ray luminosities. As a result, a large range of possible $L_X$/SFR trends with metallicity exists, if the duration of star-formation episode is not accounted for. 

Perhaps unsurprisingly, our models are in best agreement with those of \cite{Fragos2013a} who also analysed a theoretical (albeit observationally-informed) stellar population synthesis model. The \textsc{StarTrack} code \citep{Belczynski2008} used in the study is a rapid population synthesis code, and hence applies rapidly-calculated analytic estimates for remnant masses and mass transfer, rather than tracking the full structure and evolution of each modelled star. The relationship shown in Figure \ref{fig:xrayevo} \citep[and in Figure 2 of ][]{Fragos2013a} is derived specifically for HMXBs and so provides a good match to our models at times 100-200\,Myr after the onset of star formation - an epoch when the number of HMXBs has largely stabilised and before the LMXB formation rate begins to contribute significantly to the integrated X-ray luminosity.

We find that even at 10Gyrs, the X-ray flux does not converge to a constant limit, due to the long lifetime of the LMXB population and the large mass range of possible accretors. Thus, we find, as \cite{Kouroump2020} also found, that to truly use X-ray luminosity as a star-formation rate indicator, a detailed analysis of the star formation history of a galaxy is required. Neither a simple constant star formation nor a simple single-age stellar population model will describe most observed galaxies, and this will affect interpretation of their X-ray emission more strongly than that in the UV-optical.

\cite{lehmer2024}, taking a different approach, analysed 10 time bin domains across five metallicites (1.5 solar,  solar, 0.1 solar and 0.05 solar) for a population of 10$^{10}$ M$_{\odot}$. In Figure \ref{fig:xlf_age}, we show the XLFs separated into three time bin ranges with the age: $\leqslant$100~Myr, between 100~Myr and 2~Gyr and $>$2~Gyr for the 13 fuducial BPASS metallicities. In the top panel the X-ray luminosity is dominated by HMXBs and shows a wide spread in magnitude by metallicity. The final panel is dominated by LMXBs as only small companions survive to these ages. Excluding Z=0.00001, this shows very little difference in magnitude between the different metallicities. While this is not a direct comparison to \cite{lehmer2024}, our results reproduce the trend in flux with time and metallicity shown in their Figure 10, i.e. a reduction in flux at later times and a narrower spread of flux with metallicity at later times. 

\begin{figure}
\centering
   \includegraphics[width=\hsize]{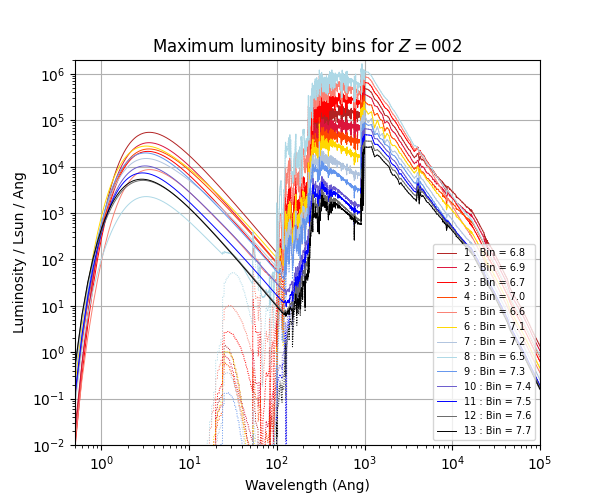}
      \caption{The spectral energy distribution for the 13 time bins with the highest peak X-ray flux for an instantaneous star burst of 10$^6$ M$_{\odot}$ of star formation for the \textsc{BPASS} metallicity $Z=0.002$. The stellar flux is shown as dotted lines and the combined stellar and X-ray flux is shown with solid lines. The ranking of the peak luminosity and the corresponding time bin is shown in the legend in units of log(age/yrs) with bin sizes of 0.1 log(age/yrs).}
         \label{fig:spec002}
\end{figure}

\begin{figure}
\centering
   \includegraphics[width=\hsize]{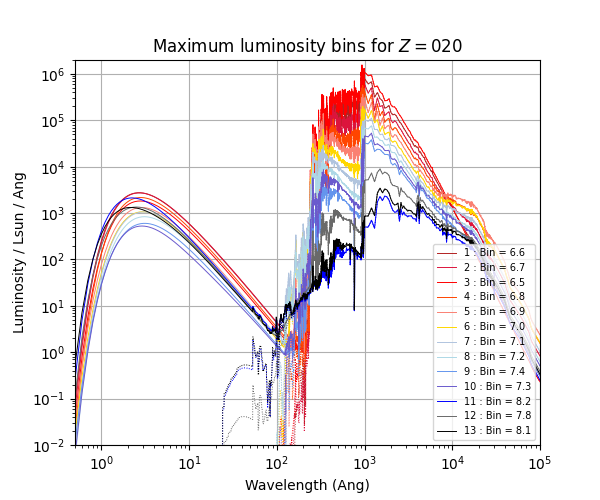}
      \caption{The spectral energy distribution for the 13 time bins with the highest peak X-ray flux for an instantaneous star burst of 10$^6$ M$_{\odot}$ of star formation for the \textsc{BPASS} metallicity $Z=0.02$. The stellar flux is shown as dotted lines and the combined stellar and X-ray flux is shown with solid lines. The ranking of the peak luminosity and the corresponding time bin is shown in the legend in units of log(age/yrs) with bin sizes of 0.1 log(age/yrs).}
         \label{fig:spec020}
\end{figure}

\section{Spectral Synthesis}\label{sec:specsynth}

\subsection{Output Spectra}

In Section \ref{sec:modval} our validation tests focussed on the number and total luminosity of XRBs in stellar populations. However, as Section \ref{sec:spectra} describes, we also calculate the accretion disc emission spectrum for each XRB in our model grid, and the integrated emission spectrum from stellar populations.

In Figures \ref{fig:spec002} and \ref{fig:spec020} we show our predicted spectra for $Z=0.002$ (1/10th solar) and $Z=0.02$ (solar metallicity) respectively, as a function of stellar population age. In both figures we show the populations with the contribution of stellar spectra alone and then with the combined stellar and X-ray emission luminosity. 

At wavelengths greater than $\sim 100\Angstrom$, the spectrum is dominated by stellar emission and there is little or no apparent change in the emission spectrum when XRB accretions discs are included in our spectral synthesis. It is only at shorter wavelengths that the XRB emission becomes apparent with the X-ray emission taking the form of a modified black body spectra, with the strongest contributions coming from the highest energy (inner) part of the accretion disc. As expected, lower metallicities have generally higher peak X-ray luminosities and their spectra tend to be slightly harder in the extreme-UV (soft X-ray) than those at higher metallicities. While the luminosity rest-UV spectra longwards of $\sim 100\Angstrom$ show a systematic trend with stellar population age (such that older populations are less UV luminous), the delay times associated with XRB formation, RLOF and mass transfer between binary components (and the dependence of all of these on metallicity) mean that this is not necessarily true in the extreme-UV to X-ray regime. At $Z=0.002$, the X-ray emission peaks at ages of around 6\,Myrs, while at $Z=0.02$ the peak is a little earlier at ages of around 4\,Myr.

\subsection{Implications for high-ionisation nebular emission lines}\label{nebemislines}

As already mentioned, one of the outstanding issues in trying to understand high redshift galaxies is the apparent excess of nebular emission from \heii\ in the UV at $1640\Angstrom$, relative to the rest-UV continuum at around 1500\,\AA\ and the hydrogen recombination lines (e.g. Ly$\alpha$ 1216\AA, H$\beta$ 4861\AA). Works including \cite{Schaerer2019} and \citet{2019A&A...621A.105S} have suggested that XRBs could resolve the issue by providing a source of ionizing photons with $h\nu>54.4$\,eV (too hot for the black body radiation associated with stellar photospheres).
However, others have suggested this is not the case. \citet[][]{Senchyna2020}, for example, considered a sample of 11 local star-forming galaxies and compared their observed HMXB point sources (observed in the X-ray) with \heii\ nebular emission (measured in the optical 4686\,\AA\ line) and its ratio to H$\beta$. They interpreted these data with models that combined \textsc{BPASS} v2.2 stellar populations with an analytic prescription for the total XRB emission estimated from the population. Their results suggested that the \heii/H$\beta$ and $L_X$/SFR ratio in the observed sources could not be reproduced simultaneously, given their model assumptions.

We have calculated how the addition of XRB emission in our models affects the number of ionizing photons from stellar populations. In Figure \ref{fig:HII} we show the time evolution in $Q(H^0)$, the number of hydrogen-ionizing photons expected for a $10^6$M$_{\odot}$ instantaneous burst of star formation. We calculate this as the number of H\,{\sc I} ionizing photons with energies greater than 13.6eV (wavelengths less than 911.8$\Angstrom$) in a given simple stellar population. We see that for all metallicities, at all times the stellar population, dominates the emission. This is expected since the number of ionizing photons is already well reproduced by typical young stellar populations.

\begin{figure}
\centering
   \includegraphics[width=\hsize]{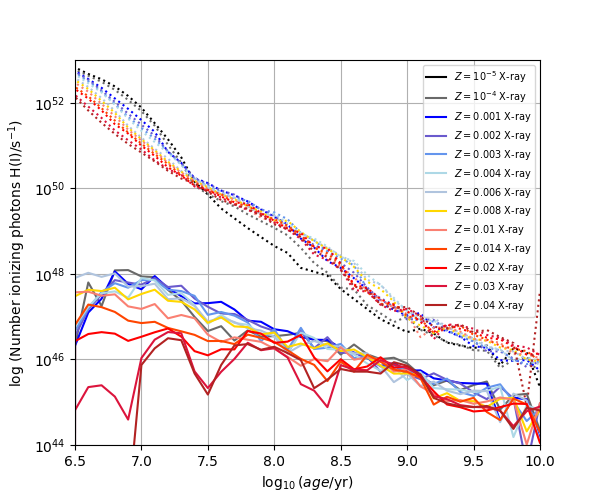}
      \caption{The number of H\,{\sc I} ionizing photons from an instantaneous burst of 10$^6$ M$_{\odot}$ of stars formed. The solid lines are for the XRB population and the dotted lines for the stellar population.}
         \label{fig:HII}
\end{figure}

\begin{figure}
\centering
   \includegraphics[width=\hsize]{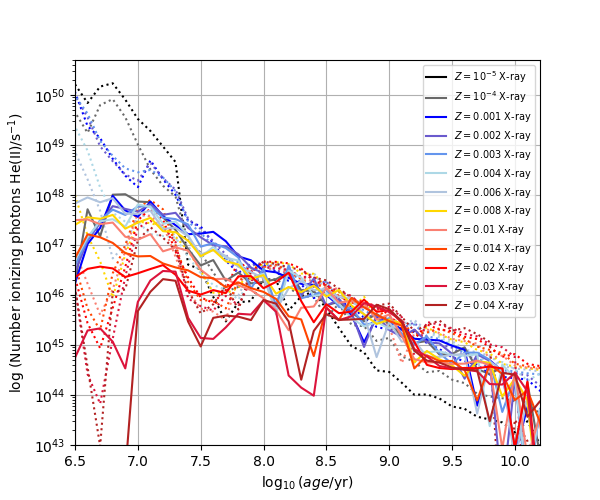}
      \caption{The number of \heii\ ionizing photons from an instantaneous burst of 10$^6$ M$_{\odot}$ of stars formed. The solid lines are for the XRB population and the dotted lines for the stellar population.}
         \label{fig:HeII}
\end{figure}

When we consider $Q(He^+)$, the number of \heii\ ionizing photons, (Figure \ref{fig:HeII}) with energies greater than 54.4eV (wavelengths less than 227.8$\Angstrom$), we can see that the contribution of photons from the stellar population and the X-ray binary population is very similar for ages beyond a few 10s of Myrs. Thus, the X-ray contribution to line strengths in the helium recombination lines cannot be neglected.

Figure \ref{fig:HII_HeII} demonstrates the time evolution of the ratio between $Q(H^0)$ and $Q(He^+)$.  A small value in this ratio indicates comparatively strong emission in the helium recombination lines relative to those of hydrogen. The smallest ratios are obtained at log(age/years)$>8$, as the stellar population ages and the contribution of LMXBs becomes increasingly significant. However, by this point, the number of hydrogen-ionizing photons from the stellar population has fallen by four orders of magnitude from its peak, and neither the hydrogen nor helium recombination line series make any significant contribution to the emission spectrum. 

A number of observations of extreme star forming galaxies both in the local Universe and high redshifts have now estimated $Q(H^0)$/$Q(He^+)$ ratios in the range $100-1000)$ \citep[e.g.][]{2013MNRAS.432.2731K,Kehrig2015,Kehrig2018,2020MNRAS.498.1638K,Berg2019,2019MNRAS.488.3492S}. 
See also discussion in \cite{2022ARA&A..60..455E}. 
These are typically associated with young stellar populations, with $\log_{10}(age/yr)$$<8$, at which the HMXB population dominates.  Our new accretion models do lower the ionizing photon production ratio relative to models considering the stellar contribution alone in this age range, although the effects are weaker than at later ages. To explore whether their impact can be sufficient to explain the observed data, without exceeding available limits on the stellar population X-ray emission, in Figure \ref{fig:senchyna} we reproduce the diagnostic parameter space from Figure 9 of \citet{Senchyna2020}, over-plotting X-ray and line ratio predictions for our models at ages of 10--100 Myr (consistent with the epoch of H$\alpha$ line emission in a binary stellar population), 10--160 Myr, 10--300Myr and 10Myr--1Gyr. Metallicities $Z=0.001-0.01$ (12+log(O/H) = 7.6--8.3), are the most appropriate for comparison to the range probed by the observed galaxy sample. Line fluxes are estimated from photon production rates using the conversion factors given in table A1 of \citet{2022ARA&A..60..455E} and $L_x$ / SFR (H$_\alpha$) is calculated by the cumulative summing of the total flux and then multiplying by the timebin duration (yr) to reflect a SFR = 1M$_\odot$ / yr .

\begin{figure}
\centering
   \includegraphics[width=\hsize]{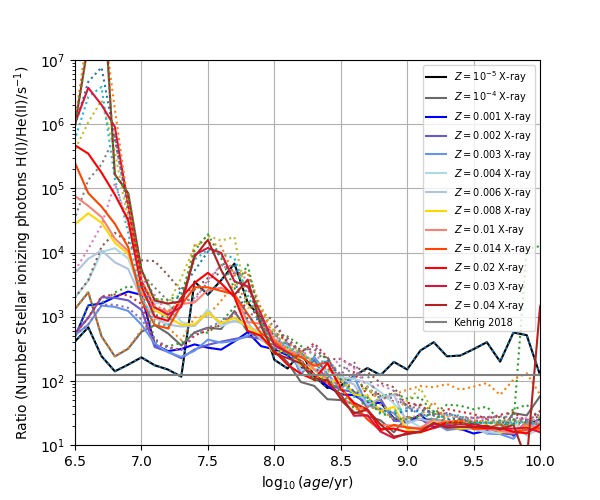}
    \caption{Ratio of H\,{\sc I} to \heii\ photons from an instantaneous burst of 10$^6$ M$_{\odot}$ of stars formed. The dotted lines are for the stellar population and the solid lines for the stellar and X-ray populations combined. The grey horizontal line is the ratio calculated from the integrated H\,{\sc I}/\heii\ photons s$^{-1}$ from \citet{Kehrig2018} for the metal-poor galaxy SBS 0335-052E.}
    \label{fig:HII_HeII}
\end{figure}

As the figure demonstrates, the new combined stellar and accretion emission models presented here venture into the parameter space which \citet{Senchyna2020} could not reach by stellar populations using their model prediction, and provides a natural explanation for at least some of the observed sources. While it falls short of explaining the strongest \heii\ lines by a factor of a few, we caution that the ionizing spectrum of massive stars (particularly at low metallicities and their $\alpha$-enhanced regimes) remains very poorly understood \citep[see discussion in][]{2022ARA&A..60..455E,2022MNRAS.512.5329B}. The emission spectrum of evolved stripped-envelope stars in these regimes also remain poorly understood, since these depend sensitively on the presumed stellar wind prescription and atmosphere models adopted. \cite{Grafener2015}, for example, have suggested that current models underestimate the narrow \heii\ wind component from such sources. 

A further caution is necessary: the models with strong helium emission generally include contributions from a moderately old (few hundred Myr) stellar population. While this is likely to contribute negligibly to the Balmer line emission or the ultraviolet continuum (which is dominated by the youngest stars), it significantly affects the mass range of accretors contributing to the X-ray flux. Such populations may not prove a good match in all the observational cases, although care should be taken to rule out the presence of contributions from an underlying, fainter stellar population which may be largely undetectable in comparison to the young stars at ultraviolet wavelengths. In Figure \ref{fig:senchyna} we show the predicted and observed HeII / H${\beta}$ line ratios using the same energy window as that used in \citet{Senchyna2020b} and \citet{Brorby2016} namely $L_{X,0.5-8keV}$/ SFR [ergs/s (M$_{\odot}$)] corresponding to H${\alpha}$ emission. 
While this figure shows that our flux is (generally), lower than that obtained by \citet{Senchyna2020} we highlight that we have limited our NS accretion rate to the Eddington limit and there is growing evidence that super-Eddington accretion is not uncommon for these objects \citep[e.g.][]{LEE2014,Gao2022,Ghodla2023}. 

While our results suggest that accreting XRBs on their own, most likely cannot explain the observed line ratios, combining the models presented here with revised stellar winds and stellar atmosphere models may well explain the emission of \heii\ completely. In addition we note from Figure \ref{fig:xlf} that our model under-predicts ULXs whose flux would certainly bring these ratios closer (as shown in Appendix \ref{a1}), and a more comprehensive treatment may bridge the gap completely. 

We continue to explore stellar atmosphere models and mass loss rates in \textsc{BPASS} in parallel projects, but such exploration lies beyond the scope of the current work.

\begin{figure}
    \centering
    \includegraphics[width=1.0\columnwidth]{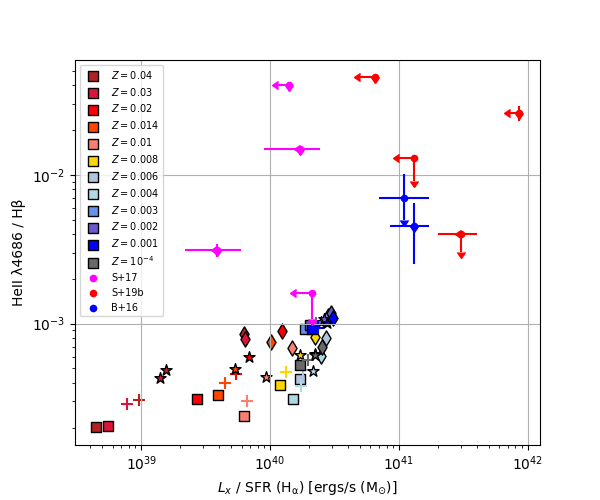}
    \caption{Predicted vs observed HeII / H$\beta$ line ratios. Observed points are from \citet{Senchyna2017} shown in magenta, \citet{Senchyna2020b} shown in blue and \citet{Brorby2016} shown in red with associated error bars. For the \textsc{BPASS} predictions we show ages 10-100My (square), 10-160Myr (star), 10-300My (plus) and 10Myr-1Gyr (diamond). The x-axis shows X-ray emission per unit SFR in the 0.5-8keV energy window, i.e. $L_{X,0.5-8keV}$ corresponding to H${\alpha}$ emission.}
    \label{fig:senchyna}
\end{figure}

Given the likely contribution of XRBs in extreme stellar populations, it is important to ask whether diagnostic emission may exist which might confirm or refute the proposed emission spectra. With the significant X-ray flux we predict, still higher ionization energy transitions may be excited and be possible to observe. For example, O\,{\sc VI} ionizing photons, those with of energies greater than 138eV (wavelengths less than 89.8$\Angstrom$) are significantly boosted in our model and are dominated at all but the very earliest ages by XRBs, as shown in Figure \ref{fig:OVI}. O\,{VI} is a challenging species to observe. Its lines at 1032 and 1038\AA\ can easily be obscured by Lyman-$\alpha$ absorption in high redshift examples (where the rest-UV is usually most straightforwardly observable), while such far-UV observations of low-redshift galaxies must be obtained from space telescopes, and are on the limits of the sensitivity range of most standard grating settings. The public data release of the CLASSY legacy survey of 45 local star forming galaxies \citep{Berg2022}, includes only four galaxies with spectral coverage of both the 1038\AA\ line and the \heii\ 1640\AA\ line. For two of these, the 1640\AA\ feature is on the noisy edge of the spectrum. There is some indication that the \heii\ line is stronger in sources with higher nebular ionization parameters (as measured in the stellar continuum-dominated [O\,{\sc III}]/[O\,{\sc II}] line ratio). However the O\,{\sc VI} line at 1037\AA\ lies between strong absorption features in C\,{\sc II} and S\,{\sc I}. It is also likely to be shock-excited, in additional to excitation by photoionization, and may also be a spectral characteristic of very massive stars ($M>100$\,M$_\odot$). Detailed analysis of this feature would require photoionization modelling of all potential sources and nearby spectral lines, which lies well beyond the scope of this paper.

\begin{figure}
\centering
   \includegraphics[width=\hsize]{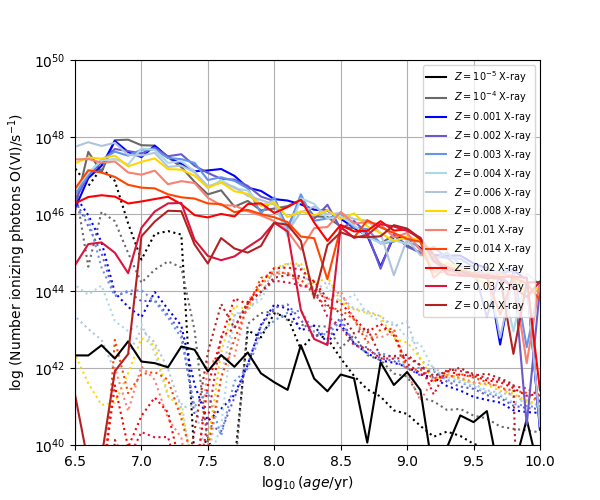}
      \caption{The number of O\,{\sc VI} ionizing photons from an instantaneous burst of 10$^6$ M$_{\odot}$ of stars formed. The solid lines are for the XRB population and the dashed lines for the stellar population.}
         \label{fig:OVI}
\end{figure}

\begin{figure}
\centering
   \includegraphics[width=\hsize]{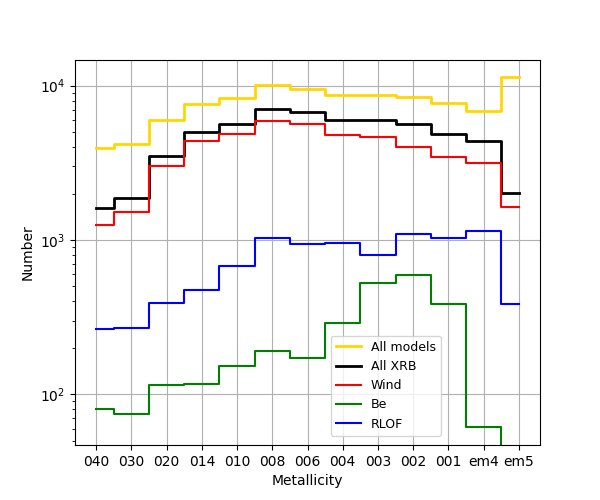}
      \caption{The total number of models making up the XRB population for each of the fuducial \textsc{BPASS} metallicities.}
         \label{fig:XRBnumbersMet}
\end{figure}

\begin{figure}
\centering
   \includegraphics[width=\hsize]{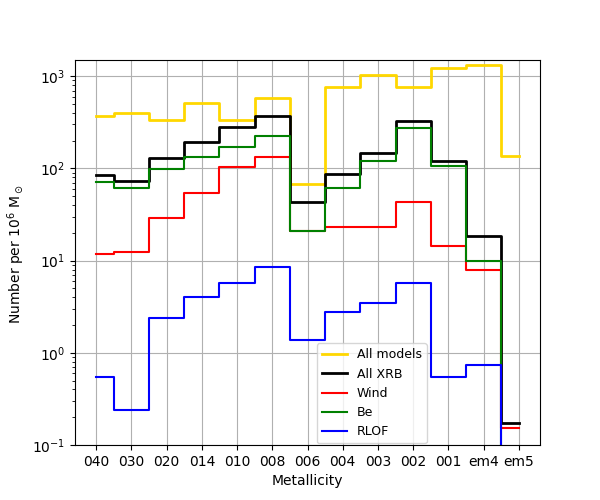}
      \caption{The total number of XRB systems in a population of $10^6$ M$_{\odot}$ for each of the fuducial \textsc{BPASS} metallicities.}
         \label{fig:XRBnumbersMetIMF}
\end{figure}
 
\begin{figure}
\centering
   \includegraphics[width=\hsize]{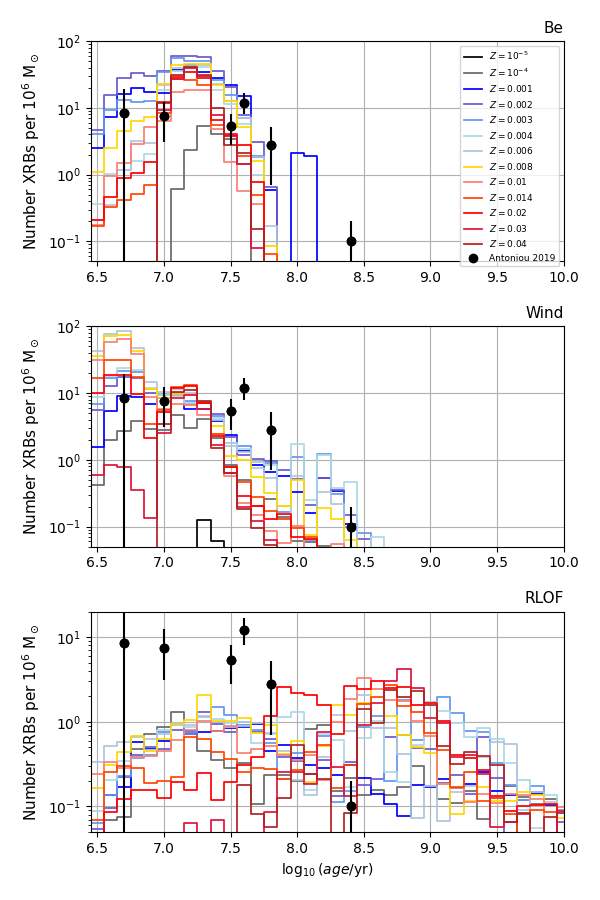}
      \caption{The total number of models making  expected in each timebin for our \textsc{BPASS} metallicities split by accrection method.}
         \label{fig:XRBnumbersTot}
\end{figure}

\section{Caveats and future work}\label{discuss}

\subsection{Mass accretion processes and luminosities}

As highlighted in Section \ref{beacc}, other than a modified accretion rate, we model Be accretion onto the CR assuming an accretion disc is formed. We then assume the same accretion disc structure and assumptions as wind-fed and RLOF processes. Due to the presence of a decretion disc, accretion onto CRs from Be stars may be more efficient than the simplistic modified \cite{Liu2023} model we use. If this were the case, the resulting increased accretion and hence luminosity of Be XRBs would extend the XLF tail shown in Figure \ref{fig:xlf} more in line with the best-fit broken power law of \cite{Mineo2011}. Any increase in X-ray flux also increases the HeII$\lambda4686$ to H$\beta$ ratio shown in Figure \ref{fig:senchyna} closer to the bulk of the observational data. 

\subsection{Uncertainties on XRB number counts}

A full synthesis of the XRB numbers and luminosities, which relies on detailed stellar interior modelling to accurately calculate the remnant masses of different binaries and the reaction of individual stellar models to mass transfer, is inevitably limited by the number of stellar models considered.  The \textsc{BPASS} v2.2 grid comprises 67 initial primary star masses in the range 0.1-300\,M$_\odot$ at each metallicity, each interacting with 189 possible secondary stars with different initial mass ratios and initial periods. However only a fraction of these (typically a few hundred) will result in the formation of accreting compact remnant binaries in any given time bin. In Figure \ref{fig:XRBnumbersMet} we show the number of \textsc{BPASS} secondary models analysed (models where a MS star has a CR companion), and then the number of models contributing to the XRB numbers by metallicity and accretion method. Approximately 10 percent of the secondary models result in XRB systems with $L_x$ > $1.38 \times$ 10$^{38}$ erg s $^{-1}$. The proportion is generally similar for each metallicity. In Figure \ref{fig:XRBnumbersMetIMF} we show the number of XRB systems for a 10$^6$ M$_{\odot}$ population. The distribution of these systems across our 50 time bins in many of our plots means bins generally have only a few data points and as a result many of the plots appear somewhat noisy.

The observed number of XRBs in a population will also be subject to significant variation due to stochastic sampling of the stellar initial mass function and the probability distribution of supernova kicks (which determine whether a binary survives and is sufficiently compact to permit transfer), especially for low-mass star clusters or low star-formation rates. These are less likely to include massive systems (e.g. HMXBs). Such concerns may be more important in the metal-enriched but low star-formation density regime of the local Universe than the galaxy-wide, low metallicity starbursts seen in distant galaxies.

At the highest metallicities stellar mass loss rates are high for massive stars, resulting in generally wider binaries as well as less massive black holes. This further decreases the number of XRBs in terms of both our predicted values ($N_{XRB}$, $L_X$) and the number of stellar models contributing to those predictions. 

In each case, the reported trends are unaffected by the Poisson uncertainties, but we caution that these modelling uncertainties on absolute numbers of XRBs and their luminosities should be taken into account in any comparison, alongside the observational selection and classification biases and the mass and age model-fitting uncertainties, on observational data. 

\subsection{Uncertainties on XRB types}

In Figure {\ref{fig:XRBnumbersTot}} we show the XRB numbers for \textsc{BPASS} metallicities split by accretion type.

Since the total XRB numbers, especially at early times, are dominated by wind-fed and Be XRB systems any uncertainties in identifying these are likely to have a significant effect on the results. In the case of Be systems, since \textsc{BPASS} does not model stellar rotation of the donor star, we have used the structure of the donor to identify Be companions to the CR (see Section \ref{beacc}), and this has generally worked well with a general trend of higher numbers of Be donors as metallicity reduces. However, we recognize that our analysis would benefit from a more robust definition for Be systems and we plan to address this in future work.

 \subsection{Uncertainties in XRB flux}
 
 While we find \textsc{BPASS} predictions in the very massive star range ($35-100$M$_{\odot}$) match well with other population synthesis codes and the observations \citep{Briel2023}, the \textsc{BPASS} flux in Figures \ref{fig:flux0402}, \ref{fig:fluxall} and \ref{fig:GILB} generally show an excess in the region $8.0 < $log(age/yrs)$ < 9.0$. This is mostly likely due to an overestimation of the number of low mass BHs forming (from progenitors below $35$M$_{\odot}$), which appear as LMXBs at these times. In addition, the \textsc{BPASS} initial mass ratio and period distributions could also influence this excess. These parameter spaces are actively being investigated by the \textsc{BPASS} group.
 
In addition, as shown in Figure \ref{fig:senchyna}, while adding the X-ray emission to the stellar emission has increased the \heii\ / H$\beta$ line ratio more in line with the observations, we see an approximate order of magnitude deficit in total X-ray flux. This is most likely a result of capping the NS luminosity at the Eddington limit. As discussed in Section \ref{nebemislines}, there is clear evidence that NSs emit well in excess of this limit, and given the very low number of NS binary models contributing to the X-ray flux, additional flux from over Eddington accretion from even a small number of NSs, would most likely eliminate this deficit. We see from  Appendix \ref{a1} a significant increase in the line ratio and this most certainly underestimates the likely increase as the post processing ignores the orbital evolution resulting from this increased mass accretion. 
  
\subsection{Future work}

The model used here for accretion disc emission has been constructed from first principles, and takes advantage of the detailed stellar modelling, remnant mass estimation and mass transfer history of the \textsc{BPASS} stellar models. However, there is naturally scope for improvement. 

\textsc{BPASS} models do not currently include a prescription for wind-driven accretion, assuming that mass transfer is dominated by Roche Lobe overflow. Current work in preparation is exploring the implementation of wind-accretion. 

We also assume that the emission from the accreting binary is dominated by the accretion disc. As already highlighted, XRB systems exist in which the primary star is irradiated by the disc of the accreting compact remnant \citep{Tout1989,Jia2016} most likely changing its evolution. Also in some XRB systems the disc may also be self-irradiated \citep{Gierliski2009,Yao2019} or irradiated by the hot compact object \citep{vanPara1994, lasota2023}, resulting in reprocessing of some fraction of the disc emission from thermal radiation into a high-energy power-law component associated with a disc wind. This will be examined in more detail in future work.

Further work will also include constraining the underlying stellar evolution models further by not only exploring the remnant mass distribution but also ensuring that the current disc models can adequately reproduce observations for individual XRBs. This may be strongly dependent on assumptions for the system duty cycles and so constrain that free parameter in our models. We note that the stellar models currently predict final remnant masses that are broadly in line with the observed masses from microlensing, gravitational wave transients and X-ray binaries \citep[see][]{Briel2023, Belczy2012}. 

The stellar population and spectral components used for comparison in these papers, are those resulting from our \textsc{BPASS} v2.2.1 stellar population and spectral synthesis with a single initial mass function \citep{StanwayEldridge2018}. Recent work in the \textsc{BPASS} project has included considering the impacts of varying the mass-dependant binary fraction and binary period and mass ratio parameters \citep{2020MNRAS.497.2201S,2020MNRAS.495.4605S}, the initial mass function \citep{2019A&A...621A.105S} and stochastic sampling of that function \citep{2023MNRAS.522.4430S}. Each of these variations can potentially impact the accretion histories and hence predicted XRB populations, although in most cases the impacts are likely to be small and we continue to recommend our observationally-motivated fiducial IMF. 

Current and planned future work is further exploring impact of $\alpha$-enhanced stellar spectra \citep[][Byrne et al in prep.]{2023MNRAS.521.4995B, 2022MNRAS.512.5329B}, and also the impact of assumptions regarding the hot star spectra (which are very uncertain, particularly at low metallicities) and of wind mass loss rates in the Very Massive Star regime \citep[see e.g.][]{2024arXiv240116165U,2022A&A...659A.163M}. While changes in these assumptions are unlikely to affect the compact remnant population, they have the potential to modify the ionizing photon production ratios between XRBs and stellar emission, particularly for high energy transitions. In addition, work on evolving the secondary star to include the mass-loss history of the primary is underway. This will provide more accurate secondary evolution models which may further affect both the remnant population and ionizing photon production.

\section{Conclusions}\label{conclude}
In this paper we have demonstrated that detailed binary stellar population models that include an XRB population can reproduce the observed X-ray luminosities of stellar populations with age and with metallicity, using realistic values for the most significant variable, the inner truncation radius of the accretion disc. Furthermore, using such complete models will be key to understanding the observations themselves - not only the integrated light of the population but also every individual binary contributing to that population.

One significant implication of our model is that we predict that XRBs can contribute to \heii\ emission in galaxies as well as that of higher ionization nebular species, without causing an excess in the hydrogen-ionizing photon flux or X-ray luminosity of the population. However, there is room for other possible sources to also contribute to the \heii\ line and XRBs are likely to be only part of a more complex picture. 

Our principal conclusions can be summarised as follows:

1) It is clear from the results presented in this paper that any meaningful analysis of X-ray emission must include binary models and their interractions otherwise key sources of high energy flux will be omitted.

2) Using first principle formulas for the compact remnant accretion rate and resulting spectra, based on mass transfer histories from detailed stellar models, we are able to reproduce the observations for the X-ray flux in M51 of \cite{Lehmer2017}

3) While we are able to manipulate the X-ray output by changing the inner accretion disc truncation radius ($R_{\textrm{inner}}$) for each compact remnant type, we reproduce the observed X-ray flux in M51 of \cite{Lehmer2017} using values in line with accepted theoretical inner truncation radius values: (WDs : $R_{\textrm{inner}}$ = 1$R_{\textrm{wd}}$, NSs : $R_{\textrm{inner}}$ = 8$R_{\textrm{s}}$ and BHs : $R_{\textrm{inner}}$ = 3$R_{\textrm{s}}$).

4) Using the same $R_{\textrm{inner}}$ values, we are able to reproduce the observed HMXB numbers for the SMC of \cite{Antoniou2019}.

5) We have calculated the X-ray flux dependence on metallicity for a number of galactic age ranges, assuming a constant star formation rate. We find that the predicted X-ray luminosity of a stellar population is highly dependent on the duration of the ongoing (and by implication any past) star-formation episode, indicating that accurate star formation history and rate information are required to estimate a meaningful $L_X$/SFR ratio. 

6) We find that while the inclusion of XRBs provides significant additional ionising photons in the \heii\ and O\,{\sc VI} energy ranges, this still cannot reproduce the most extreme \heii\ to H\,{\sc I} ratios seen in the galactic population. This is consistent with the results found by \cite{Lecroq2023} and suggests there are most likely other contributing processes in extreme starbursts. However, future work, including a more detailed analysis of the over-Eddington NS accretion and disc reprocessing, is expected to reduce this gap further.

7) While the \textsc{BPASS} models show a clear dependence of XRB flux on metallicity, the trend in the observational data is not as clear cut. Uncertainties in metallicity calculations in particular, are most likely confounding the exact observational relationship between these variables. 

8) We note that while our model somewhat over-predicts LMXB numbers when compared with some observations, this is most likely due to two factors. Firstly, we suspect that our treatment of common envelope evolution leads to fewer mergers and more surviving binary systems than seen in other models. These may accrete with a lower radiative efficiency or duty cycle than we currently assume. Secondly, our models do not consider the inevitable dynamical interactions and subsequent mergers of binary components in dense cluster environments, again resulting in more surviving binaries in our model grid and a possible over-prediction of LMXBs.

The data presented here validate the model assumptions adopted in the current generation of XRB population and spectral synthesis in \textsc{BPASS}, while suggesting some avenues for further improvement and investigation. This forms part of an ongoing effort to improve the Binary Population and Spectral Synthesis model framework and to investigate its uncertainties. This includes analysis of gravitational wave transients. While a detailed analysis lies beyond the scope of this work, we note that in addition to being able to reproduce the HMXB numbers observed in the SMC, the same \textsc{BPASS} stellar models also produce a gravitational wave transients mass spectrum which is in good agreement with the observed populations \citep[see][]{Eldridge_TANG_2018}. Such studies highlight the importance of combining information from the UV-optical-near-infrared spectrum with other electromagnetic radiation sources, and with non-electromagnetic emission where appropriate.

\section*{Acknowledgements}

JCB \& JJE acknowledge the support provided by the University of Auckland and funding from the Royal Society Te Ap$\bar{\textrm{a}}$rangi of New Zealand Marsden Grant Scheme.
ERS is supported in part by UK Science and Technology Facilities Council research grants ST/X001121/1 and ST/T000406/1.

We would like to thank the anonymous referee for their detailed and specific suggestions which have significantly improved the paper.

\section*{Data Availability}
The \textsc{BPASS} models used in this work are freely available from \texttt{https://\textsc{BPASS}.auckland.ac.nz} and \texttt{https://warwick.ac.uk/\textsc{BPASS}}.



\bibliographystyle{mnras}
\bibliography{xpass} 




\appendix

\section{Effect of super-Eddington accretion onto neutron stars}\label{a1}

As pointed out in Section \ref{bpout}, the \textsc{BPASS} models used in this work do not include super-Eddington accretion onto NSs. However, to assess the possible effects of this, we have calculated the additional mass accretion rate and luminosity assuming super-Eddington accretion onto NSs in the post-processing code. We note that this ignores the effect the additional mass accreted would have on the evolution and orbital separation for the binary system in subsequent time steps. Since, in most cases, the donor would be more massive than the NS, the most likely outcome would be that we model less evolved CRs in wider orbits with their donor stars. Nonetheless, even considering this likely underestimation, our calculations show a significant increase in X-ray flux and in the \heii\ / H$\beta$ line ratios (see Figures \ref{fig:spec020SE} and \ref{fig:senchynaSE} respectively). 


\begin{figure}
\centering
   \includegraphics[width=\hsize]{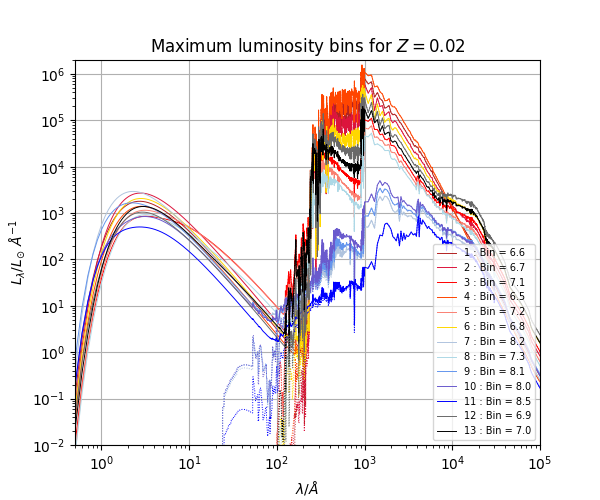}
      \caption{The spectral energy distribution for the 13 time bins with the highest peak X-ray flux for an instantaneous star burst of 10$^6$ M$_{\odot}$ of star formation for the \textsc{BPASS} metallicity $Z=0.02$. The stellar flux is shown as dotted lines and the combined stellar and X-ray flux is shown with solid lines. The ranking of the peak luminosity and the corresponding time bin is shown in the legend in units of log(age/yrs) with bin sizes of 0.1 log(age/yrs). Flux is shown with super-Eddington accretion onto NS added in the post-processing code.}
         \label{fig:spec020SE}
\end{figure}

\begin{figure}
    \centering
    \includegraphics[width=1.0\columnwidth]{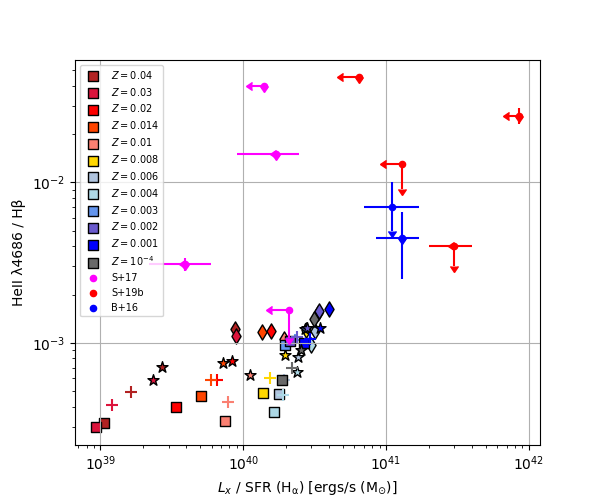}
    \caption{Predicted vs observed HeII / H$\beta$ line ratios. Observed points are from \citet{Senchyna2017} shown in magenta, \citet{Senchyna2020b} shown in blue and \citet{Brorby2016} shown in red with associated error bars. For the \textsc{BPASS} predictions we show ages 10-100My (square), 10-160Myr (star), 10-300My (plus) and 10Myr-1Gyr (diamond). Line ratios are shown with super-Eddington accretion onto NS added in the post-processing code.}
    \label{fig:senchynaSE}
\end{figure}

    \label{fig:xlfSE}


\bsp	
\label{lastpage}
\end{document}